\documentclass[aps,nofootinbib,floatfix,twocolumn,showpacs]{revtex4-1}
\usepackage{graphicx}
\usepackage{color}
\usepackage{amsmath,amssymb,bm}
\usepackage[T1]{fontenc}
\usepackage[utf8]{inputenc}
\usepackage{lmodern}
\usepackage{geometry}
\usepackage[colorlinks=true,urlcolor=blue,linkcolor=blue,citecolor=blue]{hyperref}

\begin{document}

\title{Meson structure in light-front holographic QCD}
\author{Rohit Swarnkar and Dipankar Chakrabarti}
\affiliation{Department of Physics, Indian Institute of Technology Kanpur, Kanpur 208016, India}
\date{\today}

\begin{abstract}
We consider the light-front holographic QCD with the light-front wave functions for mesons, modified for massive quarks. We evaluate the wave functions, distribution amplitudes, and form factors for $\pi$, $\rho$, $K$, and $J/\psi$ mesons and photon-to-meson transition form factors for $\pi$, $\eta$, and $\eta^\prime$ mesons. The results are compared with the experimental data, wherever available.
\end{abstract}
\maketitle


\section{Introduction}

Defining a Yang-Mills theory at the conformal $3+1$ spacetime boundary of an anti--de Sitter ($AdS_5\times S^5$) space \cite{Holographic1998} demands massless mesons. 
But it was shown that nonperturbative calculations can be performed in gauge theories with a mass gap dual to supergravity in warped spacetimes \cite{Polchinski2002}. 
One can truncate the AdS space (hard-wall) or introduce a potential in the AdS space (soft-wall) to break the conformal invariance to build confinement at large distances, while retaining conformal behavior at short distances. The mass spectrum of hadrons was computed by T\'eramond and Brodsky \cite{Teramond2005} by truncating the AdS space (in its holographic coordinate $z$) with a hard-wall set by the QCD scale ($\Lambda_{QCD}$). 

The correspondence between light-front holographic QCD and $AdS_5$ space was further utilized when hadronic wave functions were obtained by mapping an impact variable $\zeta$, which represents the measure of transverse separation of the constituents within the hadrons, to the holographic variable $z$ \cite{Brodsky2006}. Four-dimensional Schr\"odinger equations were obtained for the bound states of massless quarks and form factors were evaluated for spacelike $Q^2$.

The hard-wall model is the simplest but not the best way to incorporate confinement. The conformal invariance can be broken by introducing a dilaton background in the theory. It was found that by choosing a specific profile for a nonconstant dilaton, the usual Regge dependence can be obtained \cite{Karch2006}. The dilaton field $\varphi(z)$ produces an effective confining potential in the $AdS$ action. Choosing $\varphi(z)=\kappa^2z^2$, usual Regge behavior was observed and meson spectrum and properties have been studied within holographic QCD frameworks by several groups \cite{Brodsky2008,Forkel2007,Vega2008,Colangelo2008}. AdS/QCD wave functions and distribution amplitudes have been used to explain rho meson electroproduction \cite{Forshaw2012} and radiative $B$ decays \cite{Ahmady2013}.

The AdS/QCD model consists of massless quarks due to lack of a corresponding field in $AdS$ action which could generate quark masses. A prescription was suggested by Brodsky and T\'eramond \cite{Brodsky2008a} to include quark masses, which was further developed into models to obtain meson wave function with massive quarks \cite{Vega2009, Vega2009a}. In this work, the model developed in Ref. \cite{Vega2009a}, which incorporates massive quarks, is used to look at wave functions and distribution amplitudes for mesons. Then, form factors and transition form factors are computed with massive quarks and compared with experimental data. The model is also studied for massless quarks and compared with the massive one.

The paper is organized as follows. Original AdS/QCD model for massless quarks is discussed briefly in Sec. \ref{sec_massless}. Then, massive quarks are introduced and developed upon in Sec. \ref{sec_massive}. Wave functions and distribution amplitudes are studied in Sec. \ref{sec_wavefn_distro} for $\pi$, $\rho$, $K$, and $J/\psi$ mesons. Then we calculate form factors for the four mesons and charge radii for pion and kaon and compare the form factors and charge radii for pion and kaon with the experimental data in Sec. \ref{sec_ff}. Further, transition form factors are computed for photon-to-meson decay of $\pi$, $\eta$, and $\eta^\prime$ and plotted against experimental data in Sec. \ref{sec_tff}. Finally, $\chi^2$ per degree of freedom is calculated to quantitatively compare the predictions of the massive and the massless quark models in Sec. \ref{sec_chi}. The work is concluded in Sec. \ref{sec_conclusions}.


\section{Meson in AdS/QCD} \label{sec_massless}

Working in the Fock basis, the meson wave equation is obtained within semiclassical approximation with all the interaction terms embedded in an effective potential $U(\zeta)$. The light-front wave equation for the meson is then obtained as \cite{Brodsky2008}
\begin{eqnarray} \label{LFWFold}
\left(-\frac{d^2}{d\zeta^2}-\frac{1-4L^2}{4\zeta^2}+U(\zeta)\right)\widetilde{\phi}(\zeta)=M^2\widetilde{\phi}(\zeta),
\end{eqnarray}
where $\zeta$ is a light-front transverse variable specifying the separation between the quark and the antiquark. Whereas, the wave equation in AdS space for meson of spin-$J$ is given by
\begin{eqnarray}
\left[-\frac{z^{3-2J}}{e^{\varphi(z)}}\partial_z \left(\frac{e^{\varphi(z)}}{z^{3-2J}}\partial_z\right)+\frac{(mR)^2}{z^2}\right]\Phi_J(z)=M^2\Phi_J(z).\nonumber\\
\end{eqnarray}
Factoring out the dilaton factor from the $AdS$ field as
\begin{eqnarray}
\Phi_J(z)=\left(\frac{R}{z}\right)^{J-3/2}e^{-\varphi(z)/2}\phi_J(z),
\end{eqnarray}
and identifying $z\to\zeta$, we get the following form for $U(\zeta)$:
\begin{eqnarray}
U(\zeta)={1\over2}\varphi '' (\zeta) + {1\over4}\varphi ' (\zeta)^2+\frac{2J-3}{2\zeta}\varphi '(\zeta).
\end{eqnarray}
Using the dilaton profile of the soft-wall model $\varphi(z)=\kappa^2z^2$, the Schr\"odinger equation for the meson is obtained as
\begin{eqnarray}
\left(-\frac{d^2}{d\zeta^2}-\frac{1-4L^2}{4\zeta^2}+\kappa^4 \zeta^2+2\kappa^2(J-1)\right)\widetilde{\phi}(\zeta)\nonumber\\=M^2\widetilde{\phi}(\zeta),
\end{eqnarray}
with the mass spectrum
\begin{eqnarray}\label{mass_spec}
M^2=4\kappa^2\left(n+{{J+L}\over2}\right),
\end{eqnarray}
where $n$ and $L$ are the radial and orbital quantum numbers.

The effective light-front wave function (LFWF) for a two-parton ground state in impact space comes out as
\begin{eqnarray}  \label{LFWFeff1}
\widetilde{\phi}_{q_1\bar{q}_2}(\zeta)\sim\sqrt{x(1-x)}e^{-\frac{1}{2}\kappa^2\zeta^2}.
\end{eqnarray}
In Eq. \eqref{LFWFeff1}, using $\zeta^2=x(1-x)\textbf{b}_\perp^2$ to replace $\zeta$, where $\textbf{b}_\perp$ is the transverse impact variable that is conjugate to the light-front relative transverse momentum coordinate $\textbf{k}_\perp$, the meson LFWF can be rewritten as (along with some constants chosen for this model \cite{Vega2009a})
\begin{eqnarray}
\widetilde{\psi}_{q_1\bar{q}_2}(x,\textbf{b}_\perp)=\frac{\kappa A}{\sqrt{\pi}}\sqrt{x(1-x)}e^{\left(-{1\over2}\kappa^2 x(1-x) \textbf{b}_\perp^2\right)},
\end{eqnarray}
with a normalization parameter $A$, which will be fixed later. Performing a Fourier transform, we get
\begin{eqnarray}  \label{LFWF_kperp}
\psi_{q_1\bar{q}_2}(x,\textbf{k}_\perp)=\frac{4\pi A}{\kappa\sqrt{x(1-x)}}e^{\left(-\frac{\textbf{k}_\perp^2}{2\kappa^2 x(1-x)} \right)}.
\end{eqnarray}

In a completely different approach, namely, a domain model of QCD \cite{Nedelko}, meson spectra and wave functions have similarities with the soft-wall AdS/QCD. It will be interesting to map these two formalisms to have a better understanding of the origin of the confining potential or the particular choice of the dilaton profile.


\subsection{Massive quark model}
\label{sec_massive}

The quark masses ($m_1$ and $m_2$) are introduced by extending the kinetic energy of massless quarks with $K_0=\frac{\textbf{k}_\perp ^2}{x(1-x)}$ to
\[
K_0\rightarrow K=K_0+\mu_{12}^2, \hspace{4mm}\mu_{12}^2=\frac{m_1^2}{x}+\frac{m_2}{1-x}.
\]
This is equivalent to the change
\[
-\frac{d^2}{d\zeta^2}\rightarrow -\frac{d^2}{d\zeta^2} + \mu_{12}^2.
\]
Finally the LFWF comes out to be
\begin{eqnarray}  \label{LFWF_main}
\psi_{q_1\bar{q}_2}(x, \textbf{k}_\perp)&=& \frac{4\pi A}{\kappa\sqrt{x(1-x)}} \nonumber \\ && \times \exp\left(-\frac{\textbf{k}_\perp^2}{2\kappa^2 x(1-x)}- \frac{\mu_{12}^2}{2\kappa^2}\right).
\end{eqnarray}
This modified wave function, which incorporates quark masses within the soft-wall framework, will be used to investigate the properties of $\pi$, $\rho$, $K$, and $J/\psi$ mesons.


\subsection{Getting the parameters}

The dilaton parameter, $\kappa$, is fixed by the Regge trajectory. Now that quark masses have been introduced, the mass spectrum in Eq. \eqref{mass_spec} will be modified too. The modified spectrum should look like \cite{Teramond2009}
\begin{eqnarray}  \label{spec}
M^2 &=& 4 \kappa^2 \left( n + \frac{L + J}{2} \right)  +  \int_{0}^{1} dx \left( \frac{m_1^2}{x} + \frac{m_2^2}{1-x} \right) \nonumber \\
&&\times f^2(x,m_1,m_2) + {\rm other ~ corrections},
\end{eqnarray}
where $f(x,m_1,m_2)$ is the wave function for the longitudinal mode, which is factored out from the meson LFWF as
\begin{eqnarray}  \label{lfwffactors}
\psi(x,\zeta,m_1,m_2)=\frac{\widetilde{\phi}(\zeta)}{\sqrt{2\pi\zeta}}e^{iM\varphi}f(x,m_1,m_2).
\end{eqnarray}
For massless quarks, $f(x)$ has been found to be $f(x)=\sqrt{x(1-x)}$, which follows from an analysis of the meson form factor \cite{Brodsky2006} and is extended for massive quarks as
\[
f(x,m_1,m_2)=Nf(x)e^{-(\mu_{12}^2/2\kappa^2)},
\]
where $N$ is normalized as $1=\int_0^1dxf^2(x,m_1,m_2)$. This and other corrections to the mass spectrum are discussed in detail in Ref. \cite{Branz2010}. Other corrections come from one-gluon exchange and hyperfine-splitting contributions to the effective meson potential $U(\zeta)$. However, eventually after calculating all the corrections, it is seen that all the additional terms do not require a change in $\kappa$ to fit the experimental spectrum for the case of light mesons \cite{Branz2010}.

For $\pi$, $\rho$, and $K$, we take $\kappa=540$ MeV \cite{Teramond2010} and for J/$\psi$, $\kappa=894$ MeV \cite{Vega2009a}, which are obtained by fitting the meson mass to its respective Regge trajectory. Since $\pi$, $\rho$, and $K$ follow a Regge trajectory different from that followed by $J/\psi$, they have different values of $\kappa$. There are some other values of $\kappa$ for the pion in the literature, e.g., $\kappa=375$ MeV \cite{Brodsky2008} and $361$ MeV \cite{Vega2009}, obtained by fitting the pion form factor, and $\kappa=432$ MeV \cite{Brodsky2011}, obtained by fitting the pion transition form factor. Here, we consider $\kappa=540$ MeV, obtained through the Regge trajectory and it is the same for $\pi,~\rho,$ and $K$.

Masses and decay constants are taken from PDG and other sources. To fix the normalization parameter $A$, we use the expression for the decay constant \cite{Brodsky_decay}
\begin{eqnarray}  \label{decay}
f_M=2\sqrt{6}\int^1_0 dx\int \frac{d^2\textbf{k}_\perp}{16\pi^3}\psi_{q_1\bar{q}_2}(x, \textbf{k}_\perp).
\end{eqnarray}
All the values of the parameters used in this work are listed in Table \ref{param}. The pion has its parameters listed for zero, current, and constituent quark masses, respectively, and the kaon has its parameters listed for current and constituent quark masses, respectively, while $\rho$ and $J/\psi$ have their parameters listed only for constituent quark masses.
\begin{table}[ht]
\begin{tabular}{| c | c | c | c | c |}
  \hline
Meson & $m_1$, $m_2$ (MeV) & $\kappa$ (MeV) & $A$ & $f_M$ (MeV)\\
  \hline
$\pi$ (u,d) & 0, 0 & 540 & 0.79 & 130.7\\
		& 5, 5 & 540 & 0.79 & 130.7\\
    & 330, 330 & 540 & 2.24 & 130.7\\
\hline
$\rho$ (u,d) & 330, 330 &540 & 2.65 & 154.7\\
\hline
\textit{K} (u,s) & 5, 95 & 540 &0.99 & 156.1\\
      & 330, 500 &540 &4.69 & 156.1\\
			\hline
J/$\psi$ ($c\bar{c}$) & 1500 & 894 & 684.4 & 277.6\\
  \hline
\end{tabular}
\caption{Parameters used in this work.}\label{param}
\end{table}


\section{Wave function and distribution amplitude}
\label{sec_wavefn_distro}

\begin{figure}
\begin{center}
\includegraphics[width=0.96\linewidth,clip]{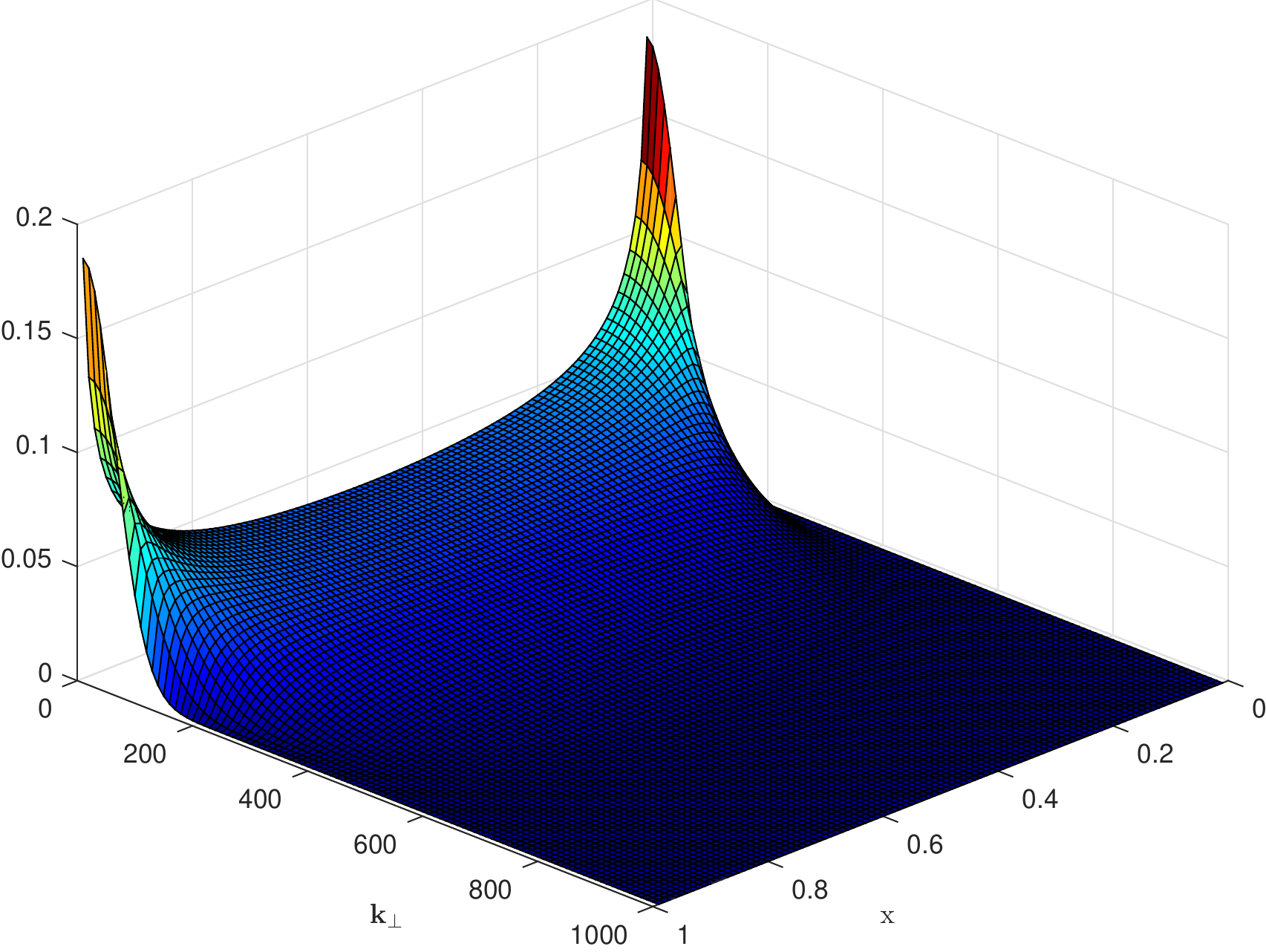} 
\caption{The pion wave function $\psi(x, \textbf{k}_\perp)$, for current quark masses $m_1=m_2=5$ MeV.}
\label{wavep_cur}
\end{center}
\end{figure}

\begin{figure}
\begin{center}
\includegraphics[width=0.96\linewidth,clip]{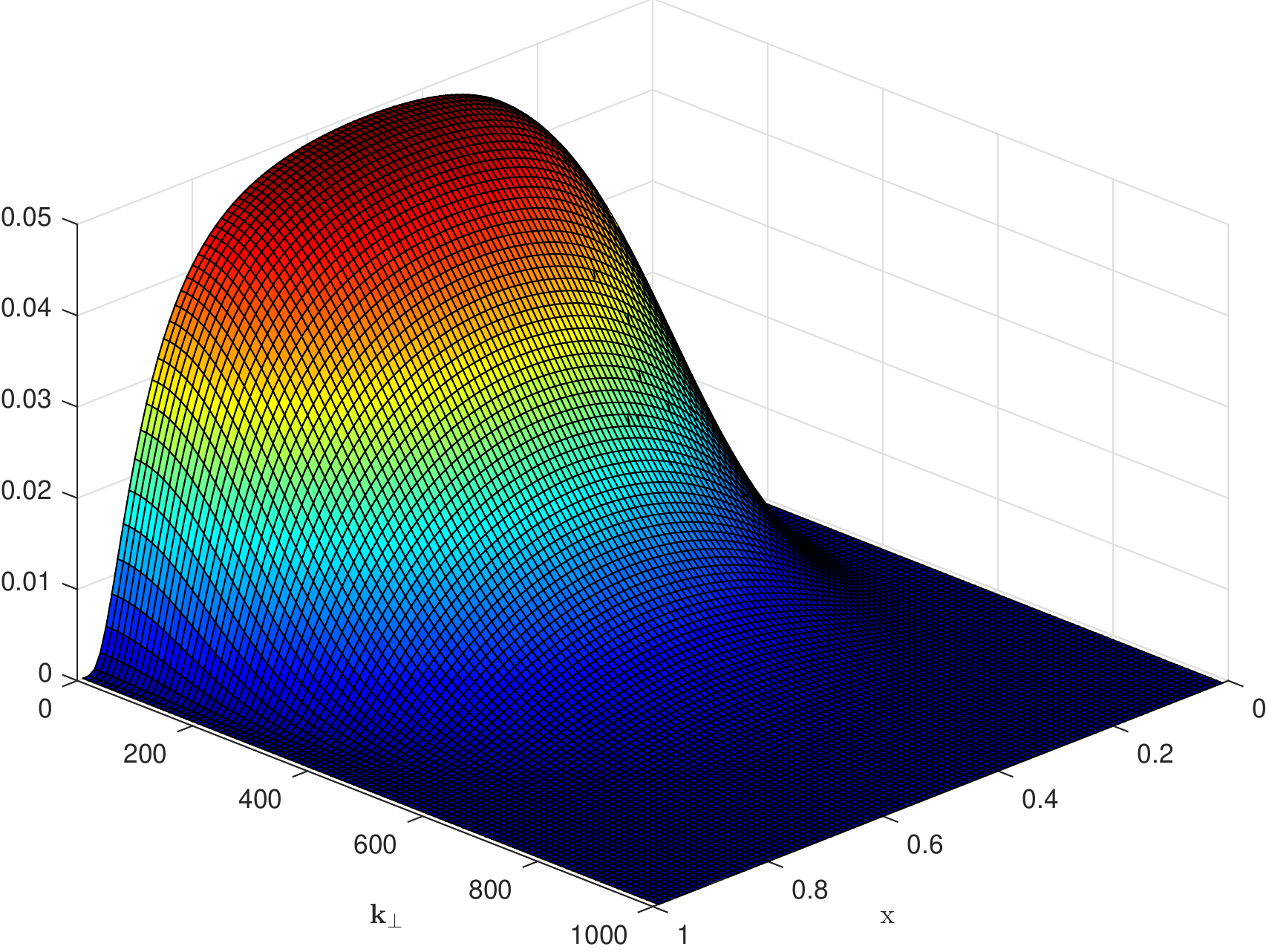} 
\caption{The pion wave function $\psi(x, \textbf{k}_\perp)$, for constituent quark masses $m_1=m_2=330$ MeV.}
\label{wavep_con}
\end{center}
\end{figure}

\begin{figure}
\begin{center}
\includegraphics[width=0.96\linewidth,clip]{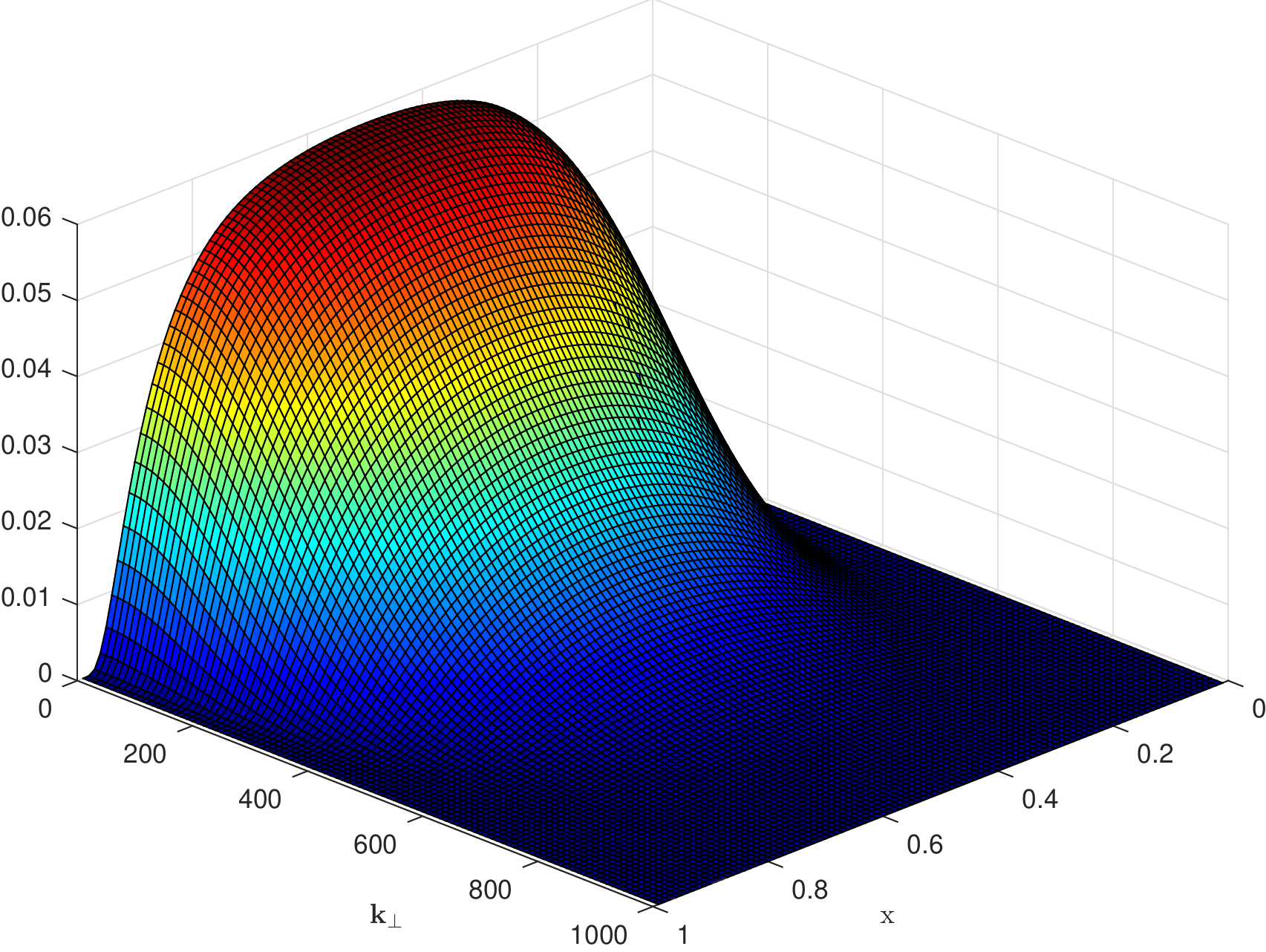} 
\caption{The $\rho$ wave function $\psi(x, \textbf{k}_\perp)$, for constituent quark masses $m_1=m_2=330$ MeV.}
\label{waver}
\end{center}
\end{figure}

\begin{figure}
\begin{center}
\includegraphics[width=0.96\linewidth,clip]{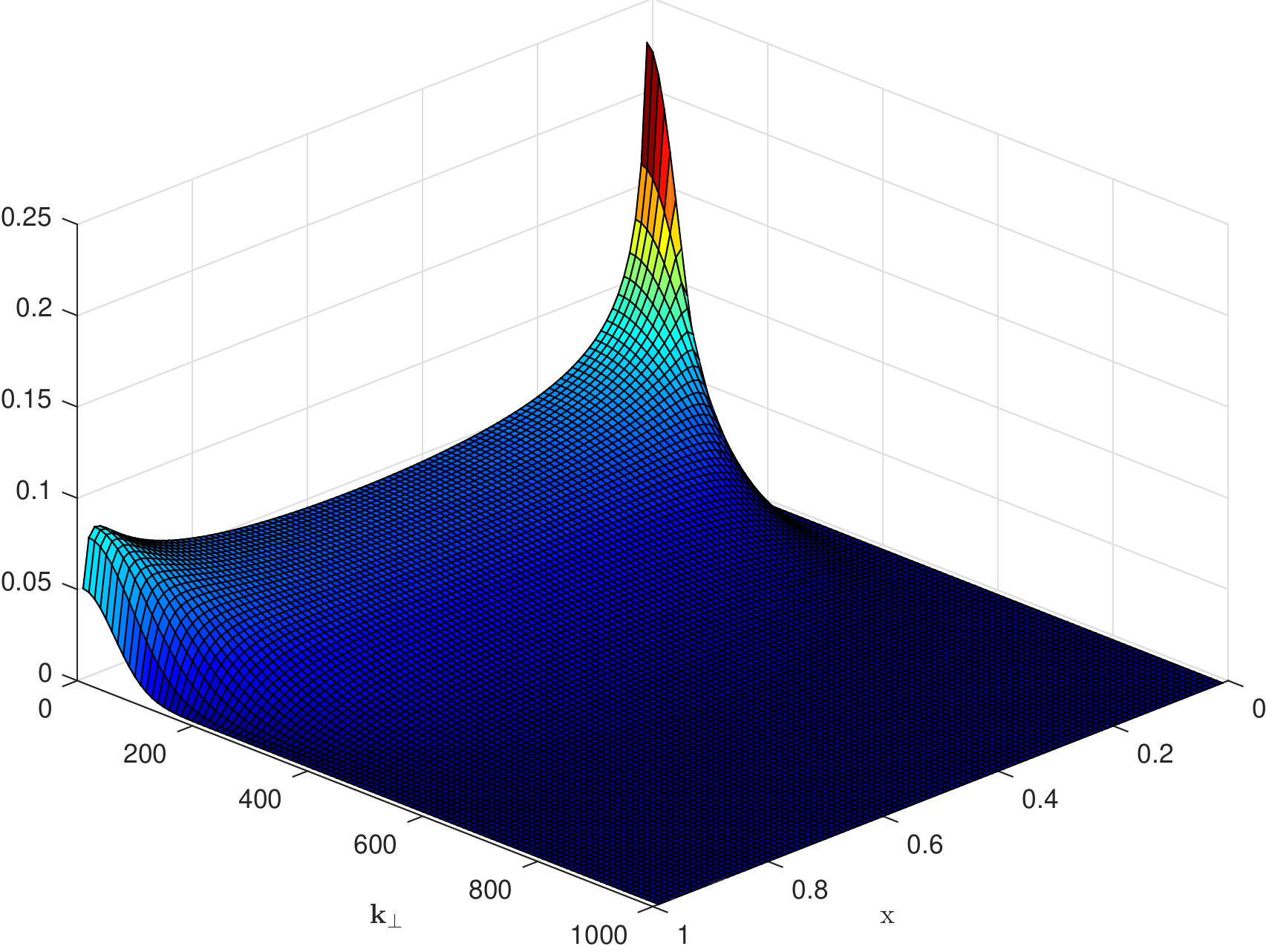} 
\caption{The kaon wave function $\psi(x, \textbf{k}_\perp)$, for current quark masses $m_1=5,m_2=95$ MeV.}
\label{wavek_cur}
\end{center}
\end{figure}

\begin{figure}
\begin{center}
\includegraphics[width=0.96\linewidth,clip]{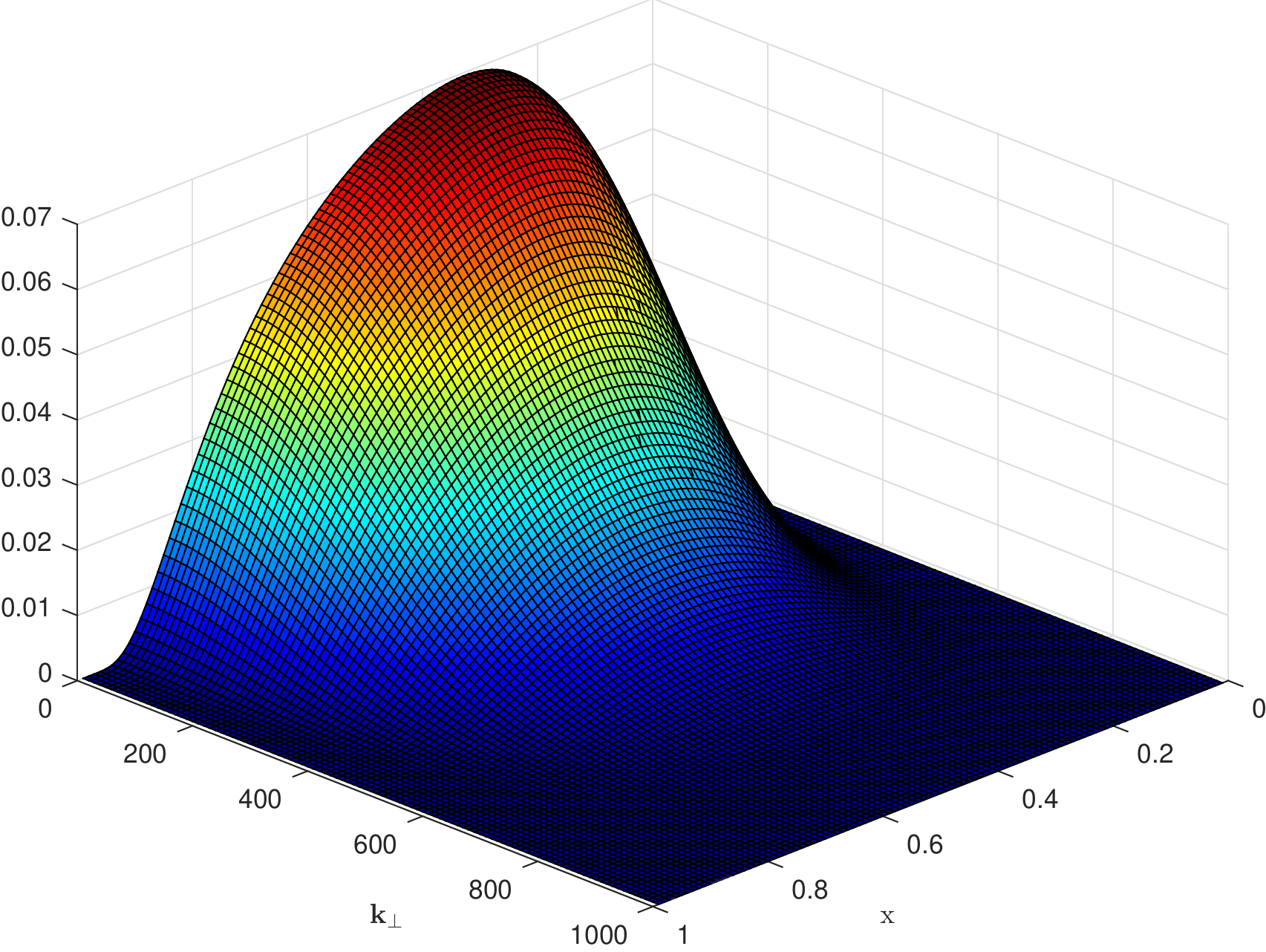} 
\caption{The kaon wave function $\psi(x, \textbf{k}_\perp)$, for constituent quark masses $m_1=330,m_2=500$ MeV.}
\label{wavek_con}
\end{center}
\end{figure}

\begin{figure}
\begin{center}
\includegraphics[width=0.96\linewidth,clip]{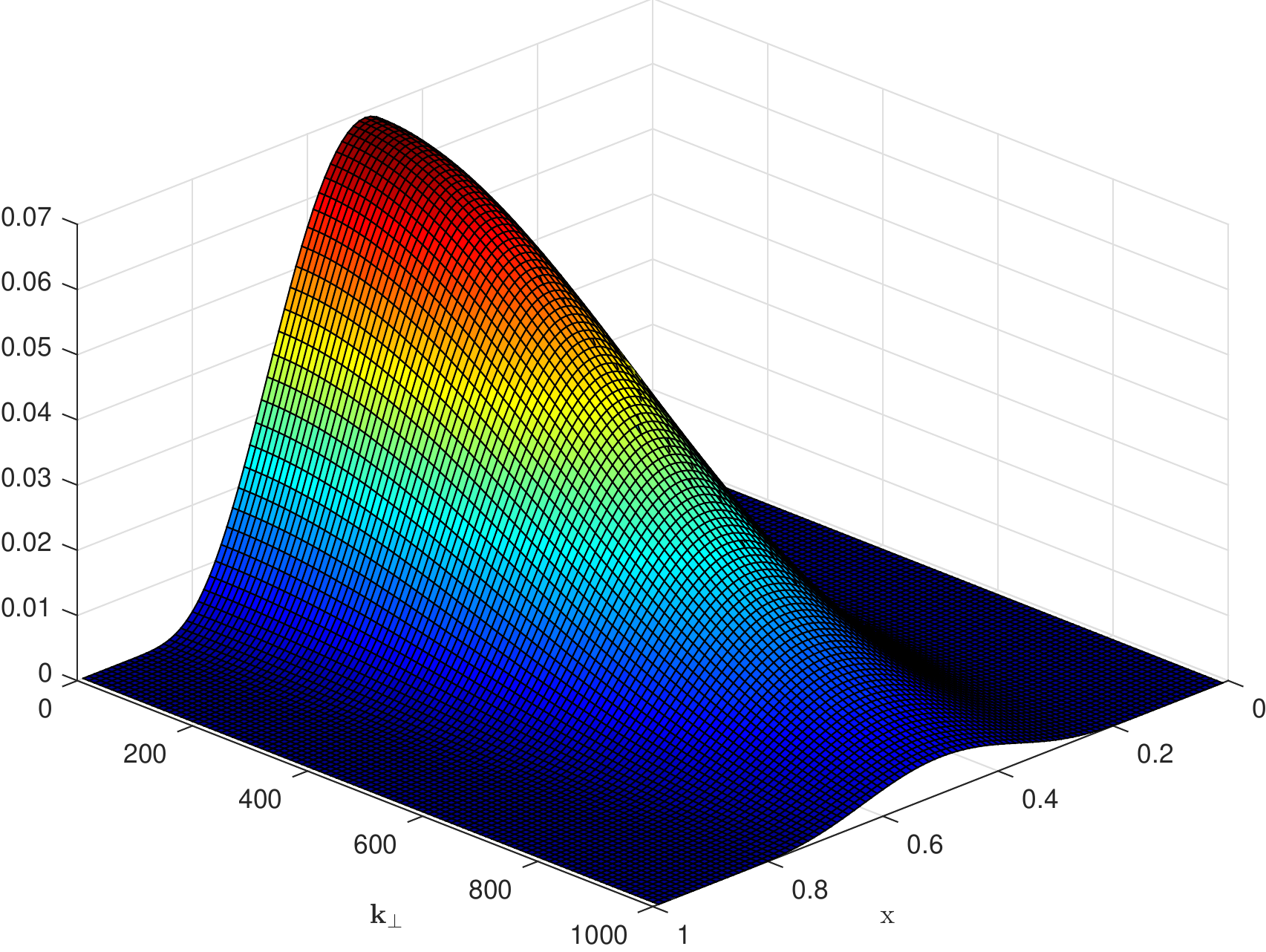} 
\caption{The J/$\psi$ wave function $\psi(x, \textbf{k}_\perp)$, for constituent quark masses $m_1=m_2=1500$ MeV.}
\label{wavej}
\end{center}
\end{figure}

For $\pi$, $\rho$, $K$, and $J/\psi$, the wave functions are plotted in Figs.~\ref{wavep_cur}-\ref{wavej} using Eq. \eqref{LFWF_main}. These graphs show the wave function narrowing down along $x$ as the mass of the meson increases. This is because of the decreasing momentum spread among the quarks for heavier mesons. For current quark masses, the wave functions peak near the end points ($x=0,1$), while for constituent quark masses, which are heavier compared to current quark masses, the wave functions peak near $x=1/2$. Also, Figs.~\ref{wavek_cur}-\ref{wavek_con} for the kaon wave function show an asymmetry in the momentum distribution because of unequal quark masses.

\begin{figure}
\begin{center}
\includegraphics[width=.96\linewidth,clip]{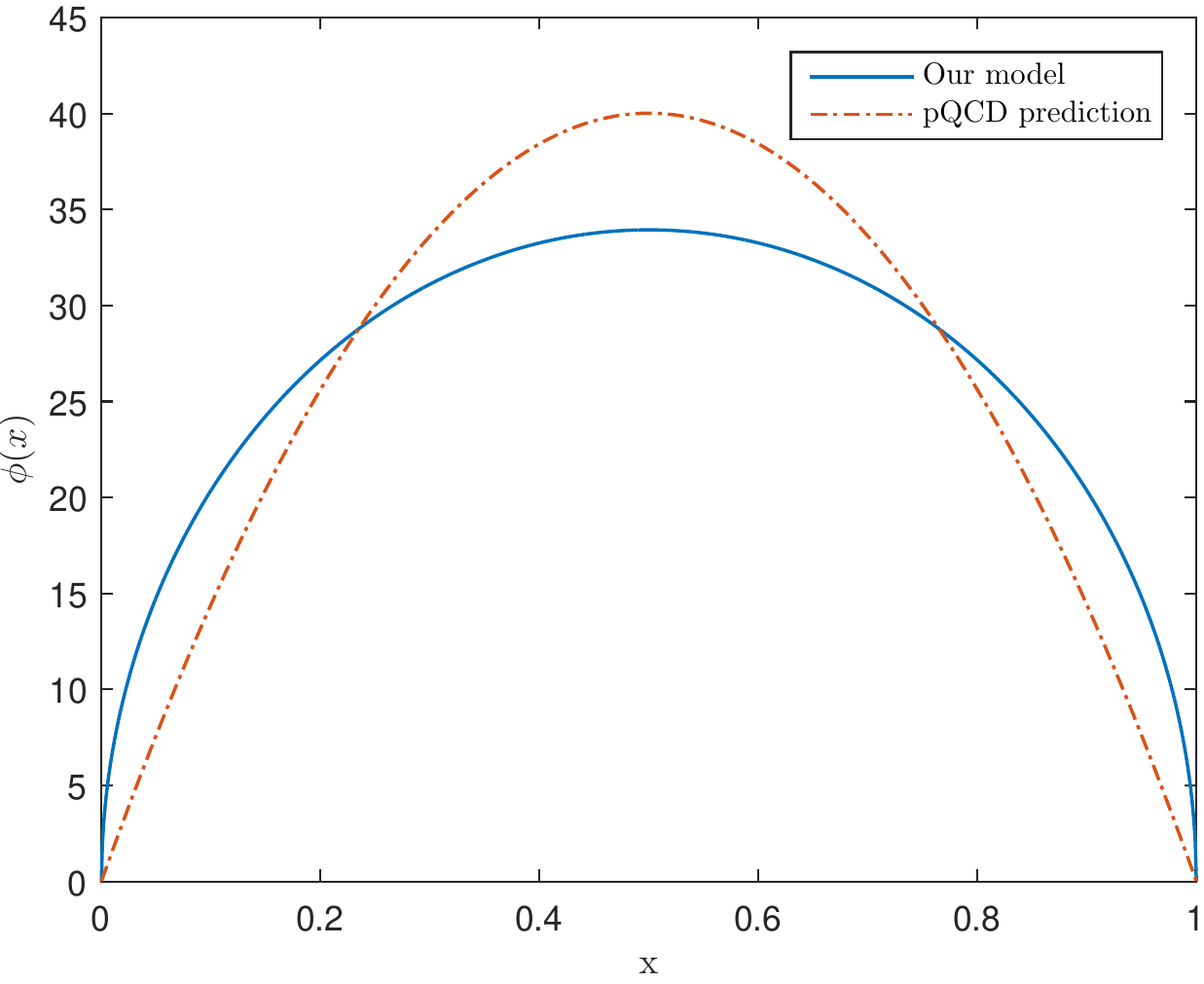} 
\caption{The pion distribution amplitude $\phi(x)$, for current quark masses, as calculated from the model (solid line) and from the pQCD prediction (dashed line).}
\label{distrop_cur}
\end{center}
\end{figure}

\begin{figure}
\begin{center}
\includegraphics[width=0.96\linewidth,clip]{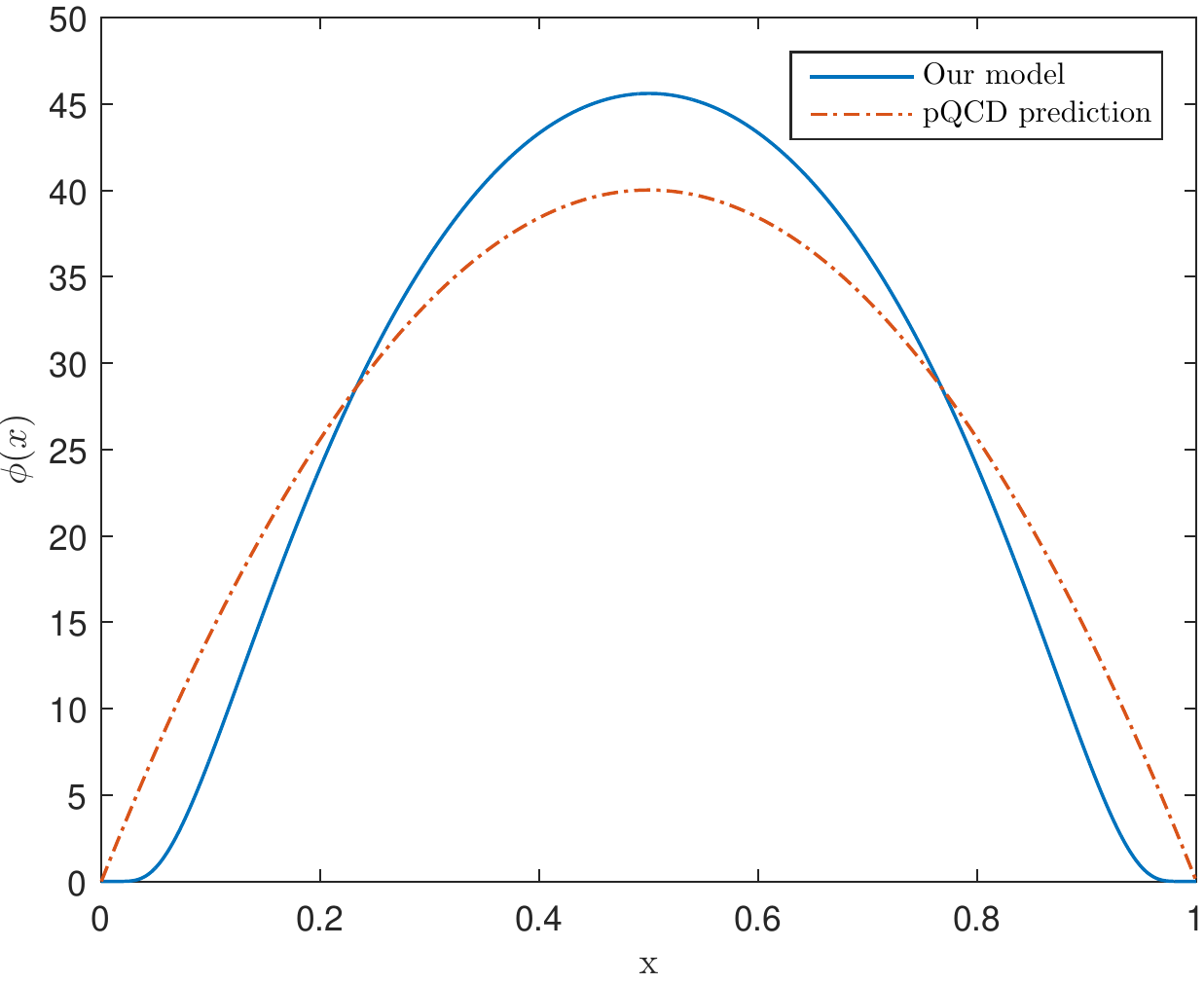} 
\caption{The pion distribution amplitude $\phi(x)$, for constituent quark masses, as calculated from the model (solid line) and from the pQCD prediction (dashed line).}
\label{distrop_con}
\end{center}
\end{figure}

\begin{figure}
\begin{center}
\includegraphics[width=0.96\linewidth,clip]{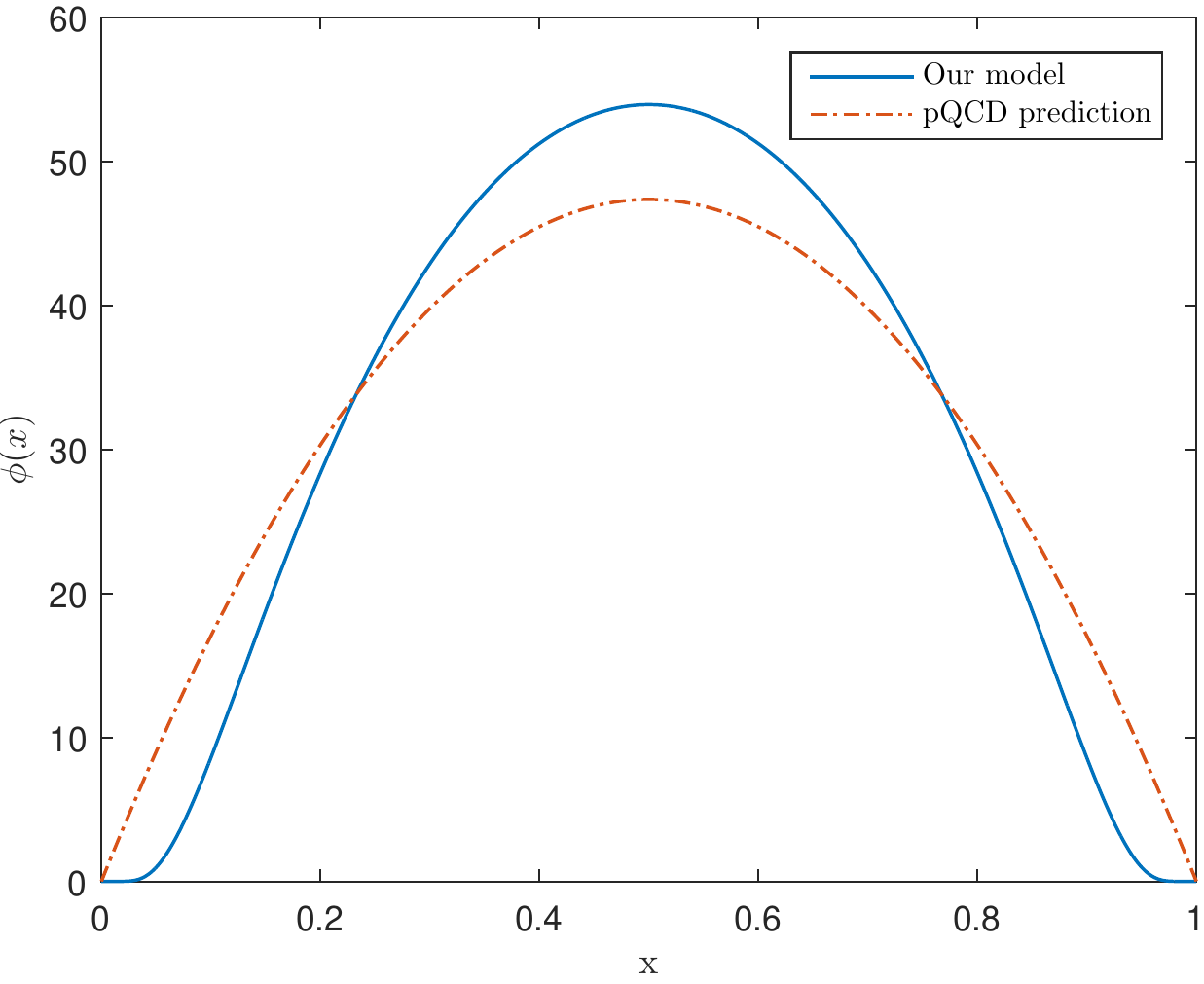} 
\caption{The $\rho$ distribution amplitude $\phi(x)$, for constituent quark masses, as calculated from the model (solid line) and from the pQCD prediction (dashed line).}
\label{distror}
\end{center}
\end{figure}

\begin{figure}
\begin{center}
\includegraphics[width=0.96\linewidth,clip]{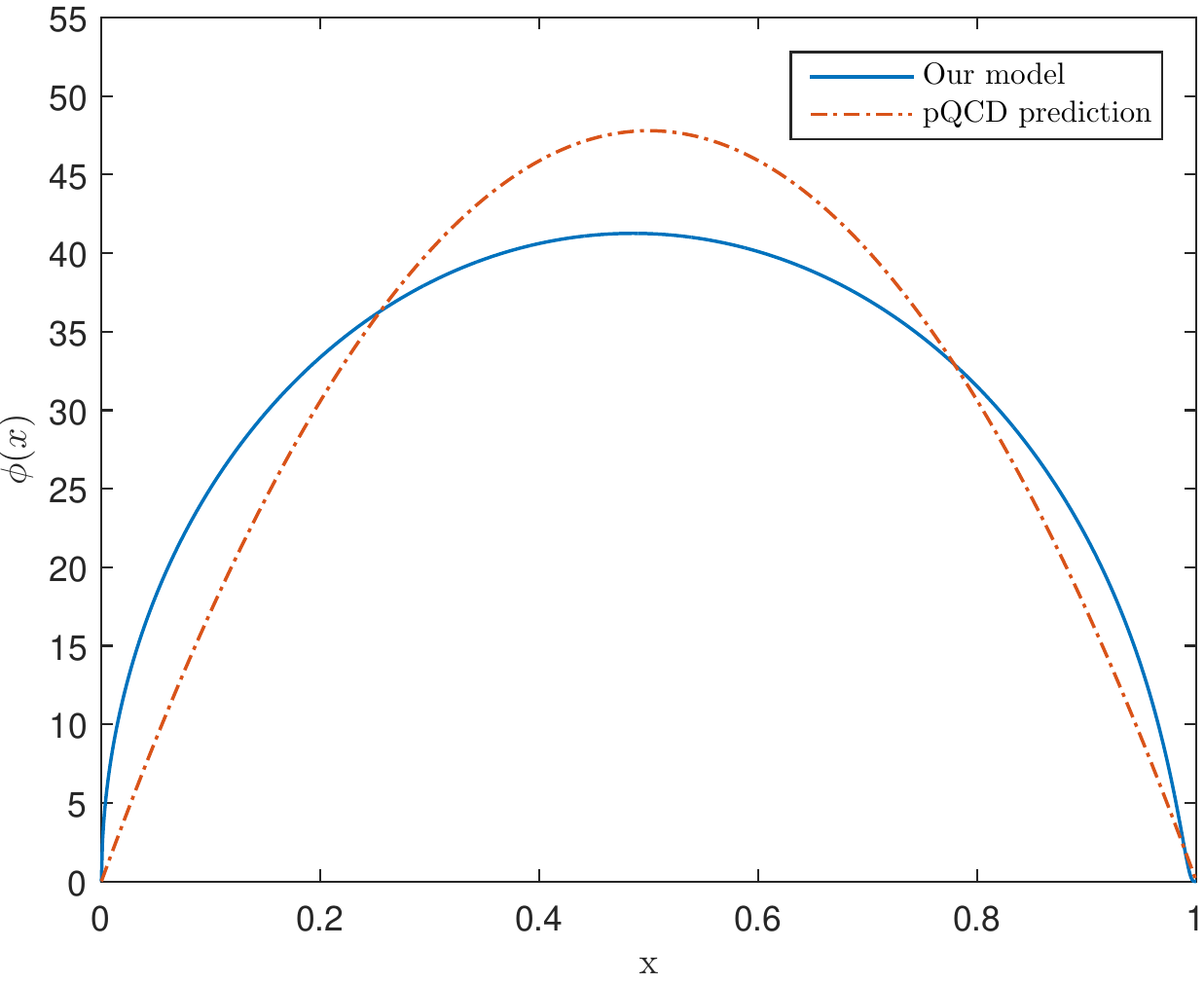} 
\caption{The kaon distribution amplitude $\phi(x)$, for current quark masses, as calculated from the model (solid line) and from the pQCD prediction (dashed line).}
\label{distrok_cur}
\end{center}
\end{figure}

\begin{figure}
\begin{center}
\includegraphics[width=0.96\linewidth,clip]{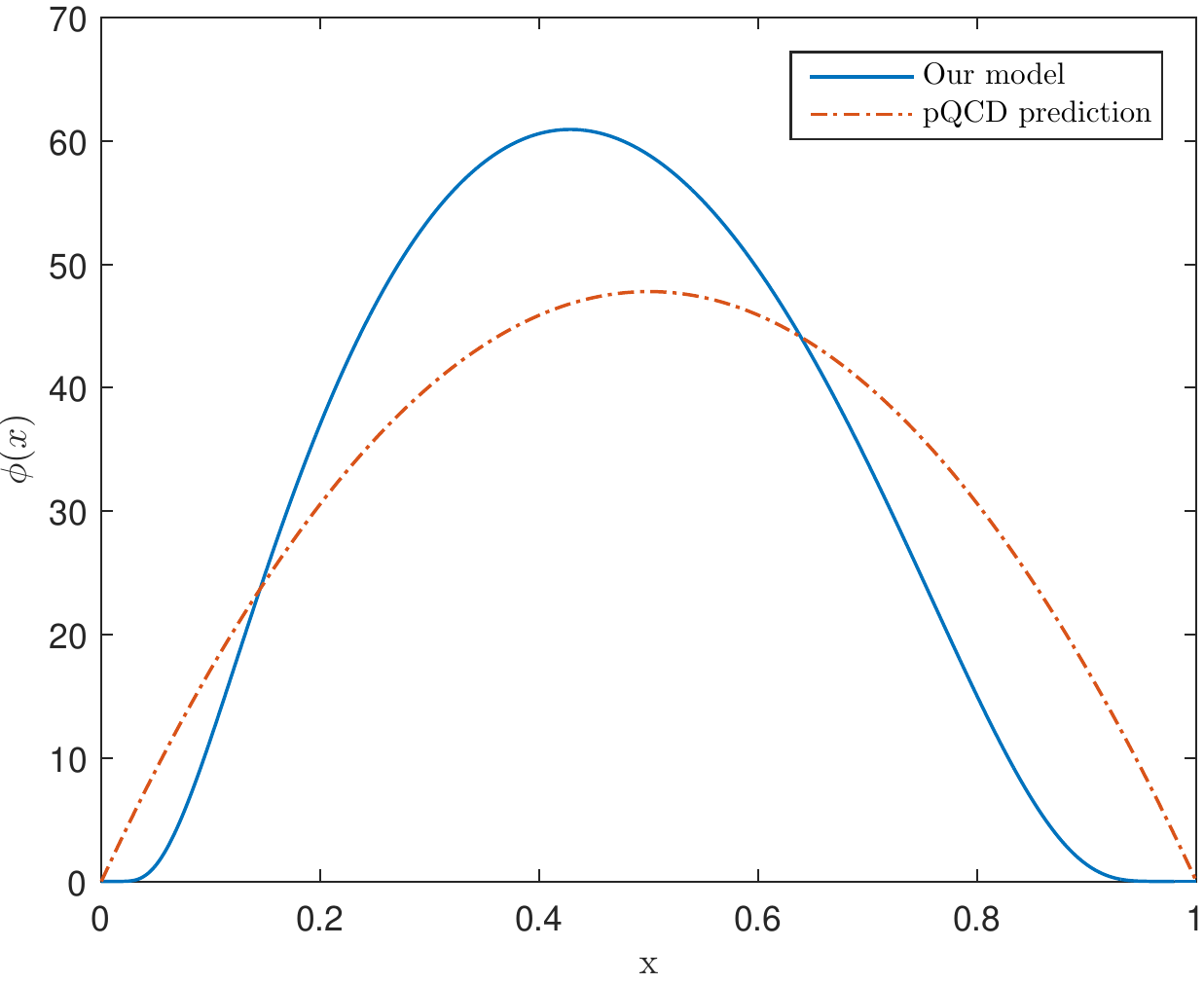} 
\caption{The kaon distribution amplitude $\phi(x)$, for constituent quark masses, as calculated from the model (solid line) and from the pQCD prediction (dashed line).}
\label{distrok_con}
\end{center}
\end{figure}

\begin{figure}
\begin{center}
\includegraphics[width=0.96\linewidth,clip]{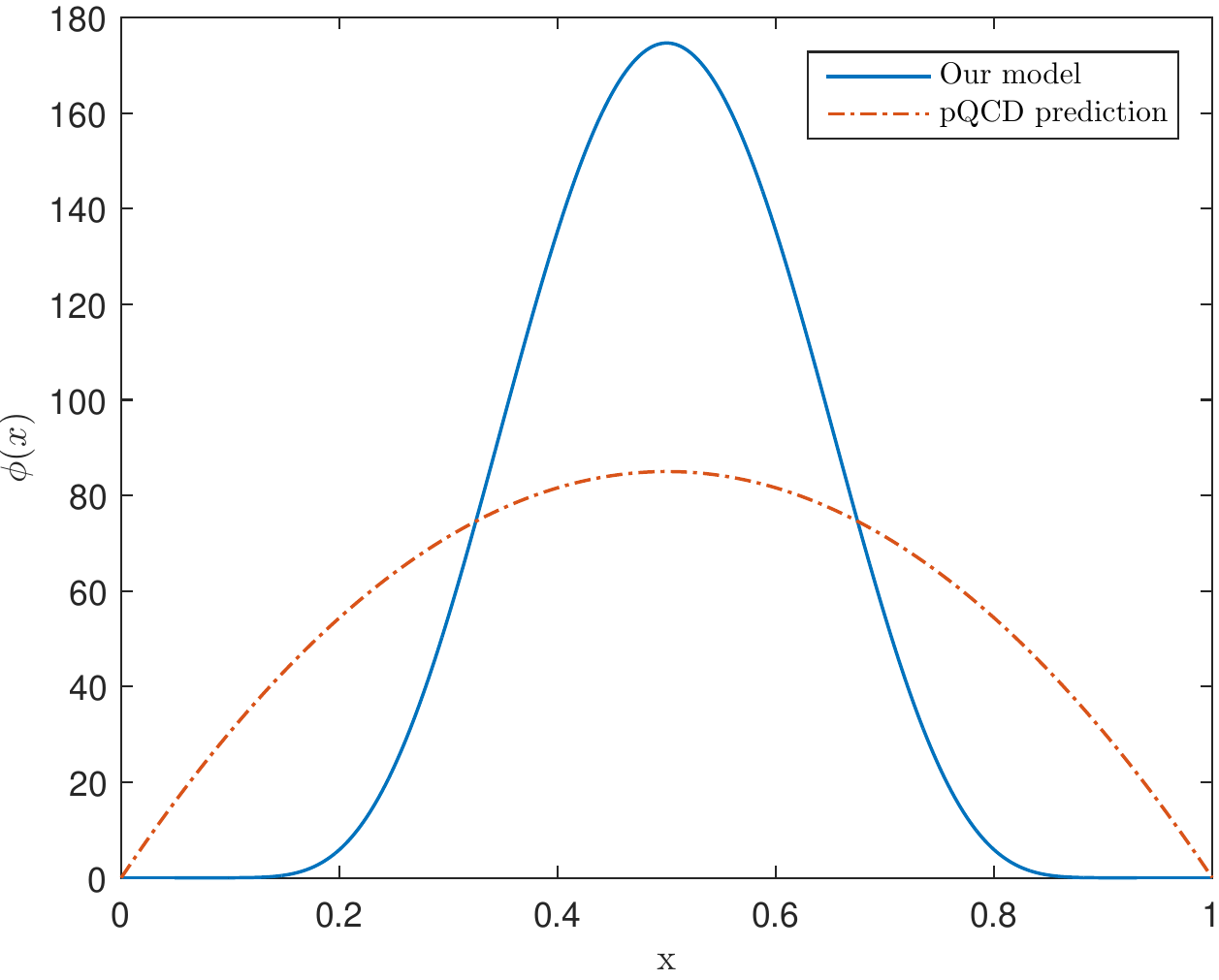} 
\caption{The J/$\psi$ distribution amplitude $\phi(x)$, for constituent quark masses, as calculated from the model (solid line) and from the pQCD prediction (dashed line).}
\label{distroj}
\end{center}
\end{figure}

The meson distribution amplitudes are important for theoretical description of electroproductions of mesons. The distribution amplitude is calculated as \cite{Lepage1980}
\begin{eqnarray}  \label{distroeq1}
\phi(x,q)=\int^{q^2}\frac{d^2\textbf{k}_\perp}{16\pi^3}\psi_{val}(x,\textbf{k}_\perp),
\end{eqnarray}
where $q^2$ is the transferred momentum squared. Although the meson $|\psi\rangle$ can be expanded into Fock states $|\psi\rangle=a_1|q\bar{q}\rangle+a_2|q\bar{q}g\rangle+a_3|q\bar{q}gg\rangle+\cdots$, for asymptotically large $q^2$, the first term dominates and we can use the wave function in Eq. \eqref{LFWF_main}, which is for the $q\bar{q}$ state, and integrate over $\textbf{k}_\perp$ to get $\phi(x)\equiv \phi(x,Q\rightarrow\infty)$ as 
\begin{eqnarray}  \label{distroeq}
\phi(x)=\frac{A\kappa}{2\pi}\sqrt{x(1-x)}\exp\left(-\frac{\mu_{12}^2}{2\kappa^2x(1-x)}\right).
\end{eqnarray}
The distribution amplitudes for $\pi$, $\rho$, $K$, and $J/\psi$ are plotted in Figs.~\ref{distrop_cur}-\ref{distroj} in which they are compared with the pQCD prediction \cite{Lepage1979} $\phi(x,q\rightarrow\infty)=\sqrt{3/2}f_M x(1-x)$. Here, again, as the meson gets more massive the distribution narrows down toward the center.

\section{Form factor and charge radius}
\label{sec_ff}

In light-front coordinates, the electromagnetic form factor can be computed from the matrix elements of the plus-component of the current $J^+$ at light-front time $x^+ = 0$ as
\begin{eqnarray} \label{Jmatrix}
\langle P' |J^+(0)|P\rangle=(P+P')^+F(q^2),
\end{eqnarray}
where $q^2 < 0$ is the transferred spacelike momentum squared. $J^+(x)$ can be computed as
\begin{eqnarray} \label{Jcurrent}
J^+(x) = \sum_q e_q\bar\psi(x) \gamma^+ \psi(x).
\end{eqnarray}
Using the momentum expansion of $\psi_+(x)=\Lambda_+\psi(x)=\gamma^0\gamma^+\psi(x)$ in terms of creation and annihilation operators \cite{Brodsky1998}
\begin{eqnarray} \label{Psiexpansion}
\psi_+(x^-,\textbf{x}_\perp)_\alpha=\sum_\lambda\int&&\frac{dq^+}{\sqrt{2q^+}}\frac{d^2\textbf{q}_\perp}{(2\pi)^3}[b_\lambda(q)u_\alpha(q,\lambda)e^{-iq.x}\nonumber\\&&+d_\lambda(q)^\dagger v_\alpha(q,\lambda)e^{iq.x}],
\end{eqnarray}
$J^+(x)$ can be expressed in the particle number representation. To compute the matrix element in Eq. \eqref{Jmatrix}, the initial and final meson states $|\psi_M(P^+,\textbf{P}_\perp)\rangle$ are expanded in terms of their Fock components
\begin{eqnarray} \label{Psifock}
|\psi(P^+,\textbf{P}_\perp)\rangle&=&\sum_{n,\lambda_i}\int\int[dx_i][d^2\textbf{k}_{\perp i}]\frac{1}{\sqrt{x_i}}\psi_n(x_i,\textbf{k}_{\perp i},\lambda_i)\nonumber\\&&\times|n:x_i P^+,x_i \textbf{P}_\perp + \textbf{k}_{\perp i},\lambda_i\rangle.
\end{eqnarray}
Then, after using a normalization condition and integrating over intermediate variables in the $q^+=0$ frame, we obtain the Drell-Yan-West expression \cite{Drell1970,West1970} for the form factor
\begin{eqnarray} \label{DYW}
F_M(q^2) = \sum_n  \int && \left[d x_i\right] \left[d^2 \textbf{k}_{\perp i}\right] \sum_j e_j \psi^*_{n/M} (x_i, \textbf{k}'_{\perp i},\lambda_i) \nonumber\\ && \times\psi_{n/M} (x_i, \textbf{k}_{\perp i},\lambda_i).
\end{eqnarray}
A Fourier transform gives the form factor in terms of $q$ as
\begin{eqnarray} \label{FT_ff} 
F_M(q^2) =  \sum_n  \prod_{j=1}^{n-1}e_j\int && d x_j d^2 \textbf{b}_{\perp j} \exp \left(i \textbf{q}_\perp \cdot \sum_{j=1}^{n-1} x_j \textbf{b}_{\perp j}\right) \nonumber\\ && \times\left|  \psi_{n/M}(x_j, \textbf{b}_{\perp j})\right|^2,
\end{eqnarray}
where $j$ runs over all $n$ constituent quarks.

For a meson, $n=2$ and there are two terms which contribute to Eq. \eqref{FT_ff}. So, we get
\begin{eqnarray}
F_M(q^2) =  \sum_{n,j} e_j \int && dx d^2\textbf{b}_\perp e^{i\textbf{q}_\perp \cdot x\textbf{b}_\perp}\left|  \psi_{n/M}(x, \textbf{b}_{\perp})\right|^2 \nonumber \\
= \sum_{n,j} e_j \int && dx d^2 \textbf{b}_\perp e^{iq|\textbf{b}_\perp|x cos\theta } \left|  \psi_{n/M}(x, \textbf{b}_{\perp})\right|^2 .\nonumber \\ 
\end{eqnarray}
Now, the integral over $\textbf{b}_\perp$ gives a Bessel function of the first kind and the sum over two quarks runs for charges $e_j\rightarrow{1\over3}, {2\over3}$ and for momentum fractions $x\rightarrow x, 1-x$ giving two integrals. Exchanging $x \leftrightarrow 1-x$ in the second integral we get 
\begin{eqnarray}  \label{ff1}
F_M(q^2)&=&2\pi\int_0^1\frac{dx}{x(1-x)} \nonumber \\ &&\times\int\zeta d\zeta J_0\left(\zeta q\sqrt{\frac{1-x}{x}}\right)\left|\psi_M(x,\zeta)\right|^2, \nonumber \\
\end{eqnarray}
where $\zeta^2=x(1-x)\textbf{b}_\perp^2$.

Now, performing a Fourier transform on our modified LFWF in Eq. \eqref{LFWF_main} and using $\zeta^2=x(1-x) \mathbf{b}_\perp^2$, we get LFWF in impact space including light-quark masses as
\begin{eqnarray}   \label{LFWFzeta}
\psi_{M}(x,\zeta) = 4\pi A\kappa \sqrt{x(1-x)}  e^{-\frac{1}{2\kappa^2}\left(\frac{m_1^2}{x}+\frac{m_{2}^2}{1-x}\right)}e^{ -{1\over2}\kappa^2 \zeta^2}.\nonumber\\
\end{eqnarray}
Using Eq. \eqref{LFWFzeta} in Eq. \eqref{ff1}, we get the meson form factor as
\begin{widetext}
\begin{eqnarray}   \label{ff2}
F_M(q^2)=(32\pi^3 A^2\kappa^2)\int^1_0dx\int\zeta d\zeta J_0\left(\zeta q\sqrt{\frac{1-x}{x}}\right)e^{-\frac{1}{\kappa^2}\left(\frac{m_1^2}{x} + \frac{m_{2}^2}{1-x}\right)}e^{-\kappa^2 \zeta^2}.
\end{eqnarray}
\end{widetext}


\subsection{$\pi$ form factor}

\begin{figure}
\begin{center}
\includegraphics[width=\linewidth]{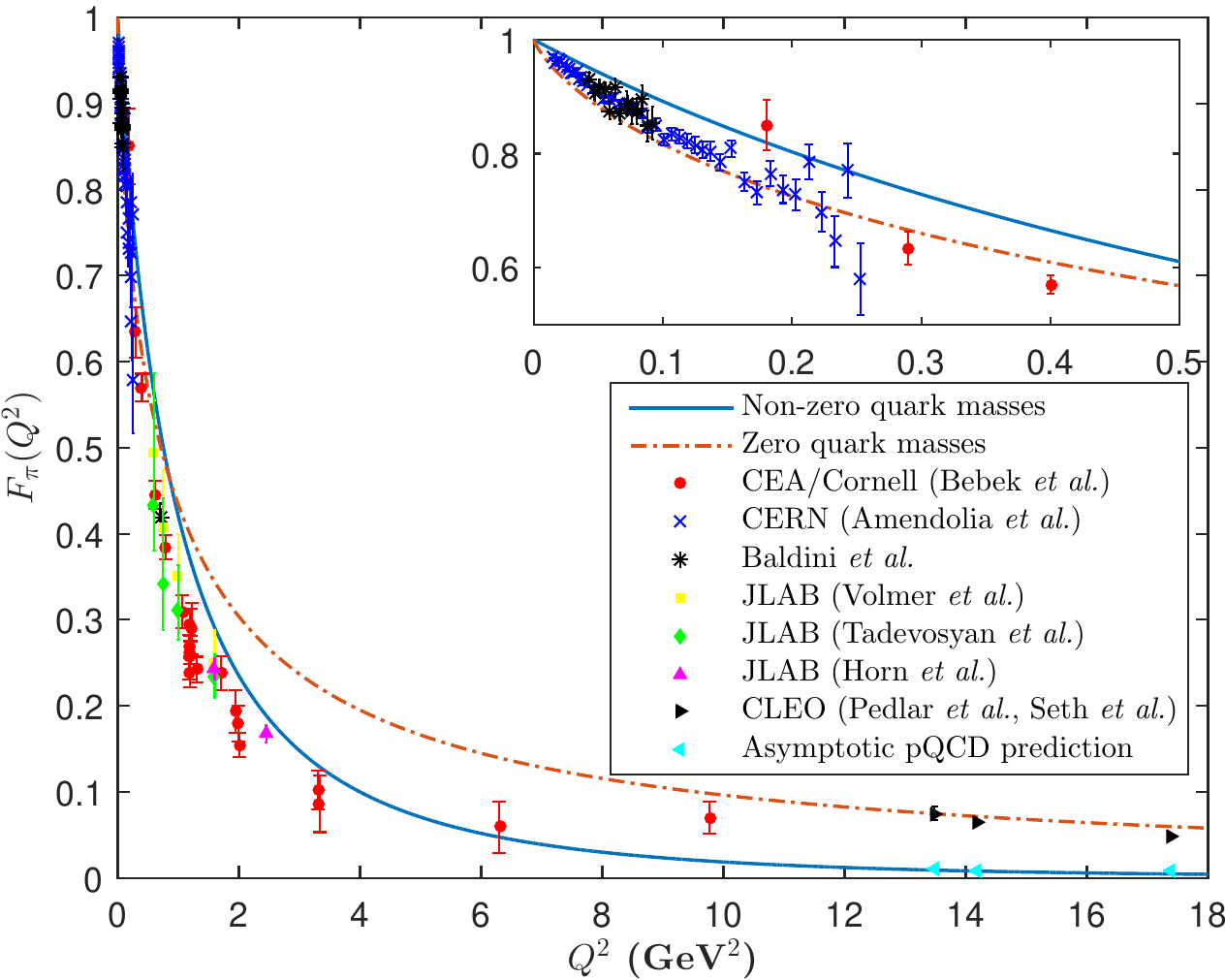} 
\caption{The pion form factor $F_\pi(Q^2)$ for nonzero (solid line) and zero (dashed line) quark masses with $\kappa=540$ MeV \cite{Teramond2010}. The plot magnified for smaller values of $Q^2$ is shown in the inset. The data are taken from \cite{Bebek1978,Amendolia1986,Baldini1999,Volmer2001,Tadevosyan2007,Horn2006,Pedlar2005,Seth2013}. The data points plotted from Baldini \textit{et al.} \cite{Baldini1999} (shown as black asterisks) are only those which do not overlap with the ones from CEA/Cornell \cite{Bebek1978} and CERN \cite{Amendolia1986}.}
\label{ff_pion}
\end{center}
\end{figure}

For the pion, the form factor $F_\pi(Q^2)$ ($Q^2=-q^2>0$) is plotted in Fig.~\ref{ff_pion} for constituent quark masses. To compare with the massless quark model, the pion form factor with zero-quark masses is also plotted. Here, $F_\pi(Q^2)$ is normalized to $F_\pi(0)=1$. The pion form factor has been measured by CEA/Cornell \cite{Bebek1978}, CERN \cite{Amendolia1986}, Baldini \textit{et al.} \cite{Baldini1999}, JLAB \cite{Volmer2001,Tadevosyan2007,Horn2006}, and CLEO \cite{Pedlar2005,Seth2013}. The data from Baldini \textit{et al.} \cite{Baldini1999} have many points which overlap with the ones from CEA/Cornell \cite{Bebek1978} and CERN \cite{Amendolia1986} and, hence, have been plotted only once. We compare our model with these experimental data sets up to $Q^2=18$ GeV$^2$ in Fig.~\ref{ff_pion}. Clearly, the pion form factor with our model fits better with the data than the one with zero-quark mass limit. For large values of $Q^2$, the asymptotic pQCD prediction for the meson form factor is \cite{Farrar1979}
\begin{eqnarray}   \label{pQCD_ff}
F_M(Q^2)=\frac{8\pi\alpha_s f^2_{M^+}}{Q^2}.
\end{eqnarray}
For $\alpha_s=0.3$ and $f_{M^+}=f_{\pi^+}=130.7$ MeV, the pion pQCD data are shown in Fig.~\ref{ff_pion}. The result from our model with massive quarks agrees well with the asymptotic pQCD prediction. This agreement is better than that with the data for large $Q^2$ values from CLEO \cite{Pedlar2005,Seth2013}.

\subsection{$\rho$ form factor}

\begin{figure}
\begin{center}
\includegraphics[width=\linewidth]{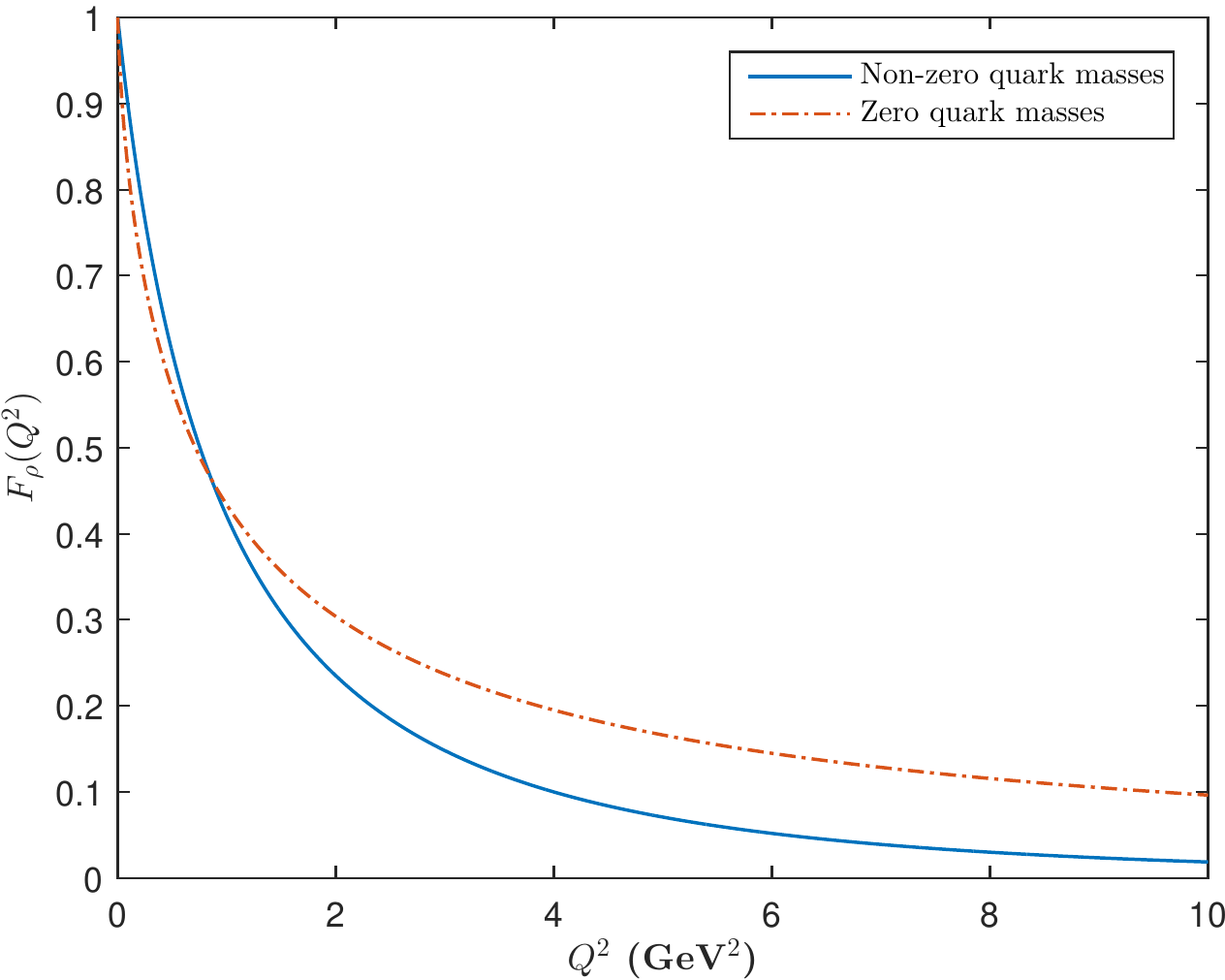} 
\caption{The $\rho$ form factor $F_\rho(Q^2)$ for nonzero (solid line) and zero (dashed line) quark masses.}
\label{ff_rho}
\end{center}
\end{figure}

For $\rho$, the form factor $F_\rho(Q^2)$ is plotted in Fig.~\ref{ff_rho}, along with the plot for zero quark masses. Here, $F_\rho(Q^2)$ is normalized to $F_\rho(0)=1$.


\subsection{$K$ form factor}

\begin{figure}
\begin{center}
\includegraphics[width=\linewidth]{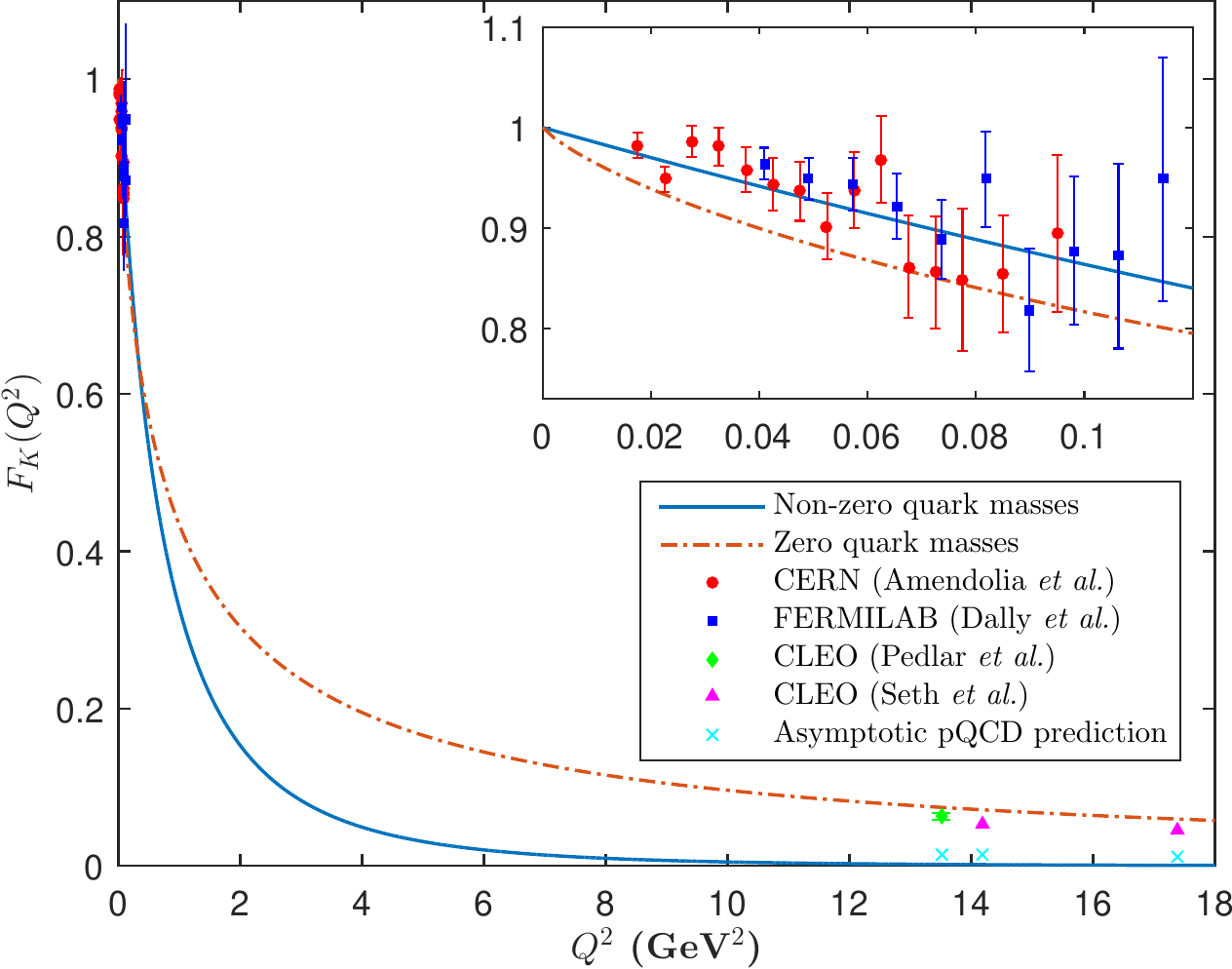} 
\caption{The kaon form factor $F_K(Q^2)$ for nonzero (solid line) and zero (dashed line) quark masses with $\kappa=894$ MeV \cite{Vega2009a}. The plot magnified for smaller values of $Q^2$ is shown in the inset. The data are taken from \cite{Amendolia1986a,Dally1980,Pedlar2005,Seth2013}.}
\label{ff_kaon}
\end{center}
\end{figure}

For the kaon, the form factor $F_K(Q^2)$ is plotted in Fig.~\ref{ff_kaon} for both nonzero and zero quark masses. Here, $F_K(Q^2)$ is normalized to $F_K(0)=1$. The kaon form factor has been measured by CERN \cite{Amendolia1986a}, FERMILAB \cite{Dally1980}, and CLEO \cite{Pedlar2005,Seth2013}. We compare our model with these experimental data sets up to $Q^2=18$ GeV$^2$ in Fig.~\ref{ff_kaon}. Similar to the pion, the values from the asymptotic pQCD prediction [Eq. \eqref{pQCD_ff}] are also plotted, which fit closely with our model.

\subsection{$J/\psi$ form factor}

\begin{figure}
\begin{center}
\includegraphics[width=\linewidth,clip]{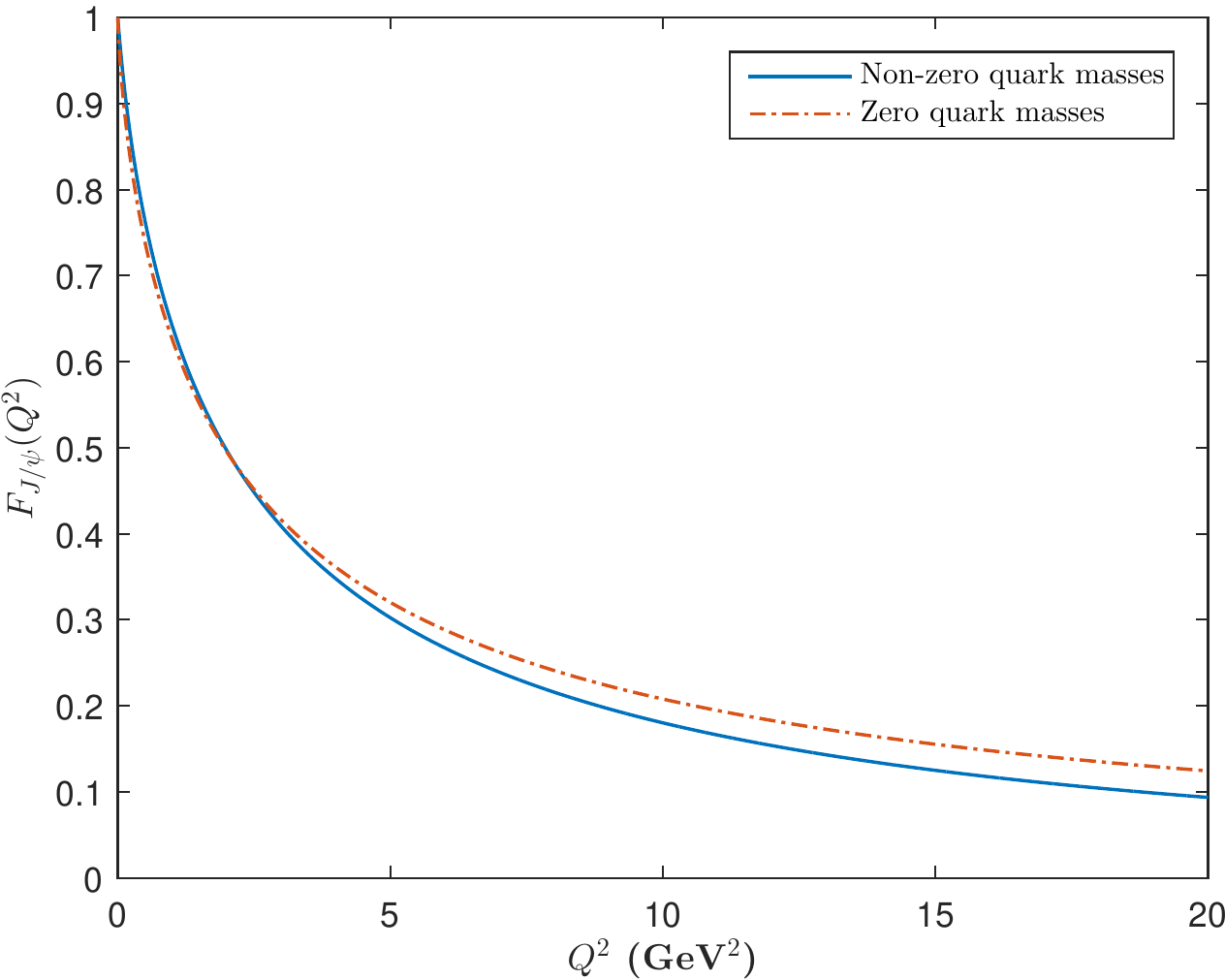} 
\caption{The J/$\psi$ form factor $F_{J/\psi}(Q^2)$ for nonzero (solid line) and zero (dashed line) quark masses.}
\label{ff_jpsi}
\end{center}
\end{figure}

For J/$\psi$, the form factor $F_{J/\psi}(Q^2)$ is plotted in Fig.~\ref{ff_jpsi}, for both nonzero and zero quark masses. Here, $F_{J/\psi}(Q^2)$ is normalized to $F_{J/\psi}(0)=1$.


\subsection{Charge radius}

The charge radius of the meson can be calculated as
\begin{eqnarray}   \label{chrg}
\langle r^2\rangle=-\frac{6}{F(0)}\left. \frac{dF(Q^2)}{dQ^2}\right|_{Q^2=0}.
\end{eqnarray}
For the pion, with $\kappa=540$ MeV, the charge radius comes out to be $\sqrt{\langle r^2_\pi\rangle} = 0.529$ fm for massive quarks and $1.679$ fm for massless quarks, while the experimental value is $\sqrt{\langle r^2_\pi\rangle} = 0.672\pm0.008$ fm \cite{PDG2014}.

For the kaon, with $\kappa=894$ MeV, it comes as $\sqrt{\langle r^2_K\rangle} = 0.598$ fm for massive quarks and $1.679$ fm for massless quarks, while the experimental value is $\sqrt{\langle r^2_K\rangle} = 0.560\pm0.031$ fm \cite{PDG2014}.

The deviation from the experimental values, significantly evident in the pion's case, can be attributed to the fact that the value of a charge radius mainly depends on the values of the form factor only at small $Q^2$ (near zero), given that we compute the slope in the $Q^2=0$ limit. In this range, our model does not agree quite well with the experimental data, which in turn causes this mismatch.


\section{Transition form factor}
\label{sec_tff}

The amplitude for two-photon production of the meson ($\gamma^*\gamma\to M$) contains an unknown function of photon virtuality ($Q^2$) which is known as the meson transition form factor. The photon-to-meson form factors are measured in several experiments \cite{Aubert2009, Gronberg1998, Behrend1991, Uehara2012, Sanchez2011}. Thus, there are many theoretical interests to predict these transition form factors. Here we present the predictions from the AdS/QCD model of massive quarks.


\subsection{$\pi$ transition form factor}

\begin{figure}
\begin{center}
\includegraphics[width=\linewidth,clip]{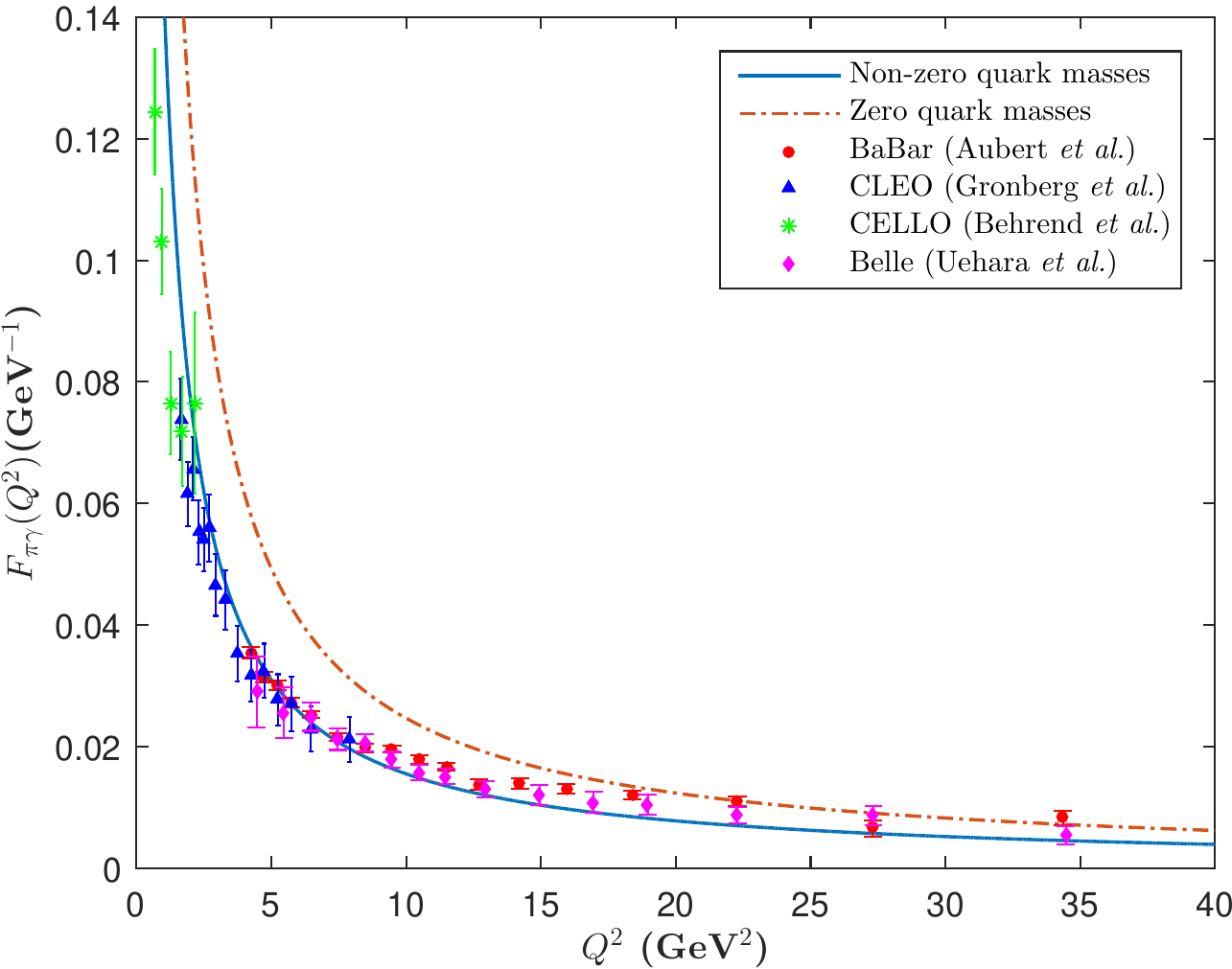} 
\caption{The pion transition form factor $F_{\pi\gamma}(Q^2)$ for nonzero (solid line) and zero (dashed line) quark masses. The data are taken from \cite{Aubert2009,Gronberg1998,Behrend1991,Uehara2012}.}
\label{tff_pion}
\end{center}
\end{figure}

\begin{figure}
\begin{center}
\includegraphics[width=\linewidth,clip]{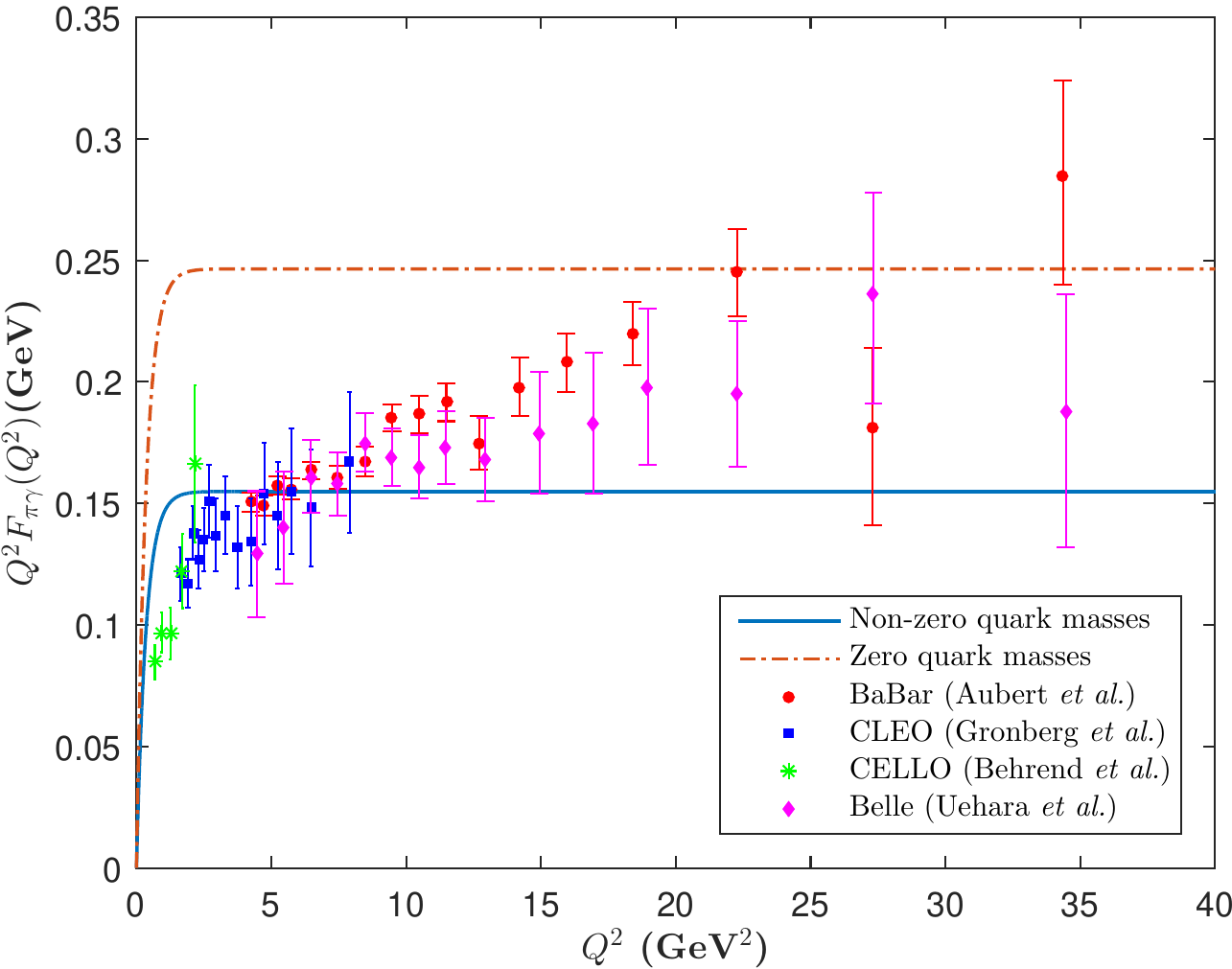} 
\caption{The pion transition form factor $Q^2F_{\pi\gamma}(Q^2)$ for nonzero (solid line) and zero (dashed line) quark masses. The data are taken from \cite{Aubert2009,Gronberg1998,Behrend1991,Uehara2012}.}
\label{qtff_pion}
\end{center}
\end{figure}

The photon-to-pion transition form factor $F_{\pi\gamma}(Q^2)$ can be calculated in QCD as \cite{Lepage1980} ($Q^2=-q^2>0$)
\begin{eqnarray}   \label{eqtff_orig}
Q^2F_{\pi\gamma}(Q^2)=\frac{4}{\sqrt{3}}\int_{0}^{1}dx\frac{\phi(x,\bar{x}Q)}{\bar{x}}\left[1+O\left(\alpha_s,\frac{m^2}{Q^2}\right)\right],\nonumber\\
\end{eqnarray}
where $x$ is the longitudinal momentum fraction of the quark hit by the virtual photon in the hard scattering process and $\bar{x}=1-x$ is for the other spectator quark. The distribution amplitude $\phi(x,Q)$ is given by Eq. \eqref{distroeq1}. In pQCD, the pion distribution amplitude for $Q^2\to\infty$ is given by $\phi(x,Q^2\to\infty)=\sqrt{3/2}f_\pi x(1-x)$. Thus, the pQCD prediction for the pion transition form factor is given by $\lim_{Q^2\to\infty} Q^2F_{\pi\gamma}(Q^2) =\sqrt{2}f_\pi$ where $f_\pi$ is the pion decay constant. The $Q^2$ dependence of the form factor helps to constrain the models for distribution amplitudes. In our model, the distribution amplitude is given by Eq. \eqref{distroeq} and the expression for the pion transition form factor can be written as
\begin{eqnarray}   \label{eqtff_pion_mid}
Q^2F_{\pi\gamma}(Q^2)&=& \frac{A}{\sqrt{3}\pi^2\kappa}\int^1_0 \frac{dx}{1-x}\frac{1}{\sqrt{x(1-x)}}\nonumber\\&&\times\int_0^{(1-x)^2Q^2}d^2\textbf{k}_\perp e^{\left(-\frac{\textbf{k}_\perp^2}{2\kappa^2x(1-x)}-\frac{\mu_{12}^2}{2\kappa^2}\right)}.\nonumber\\
\end{eqnarray}
This simplifies to the following form
\begin{widetext}
\begin{eqnarray}   \label{eqtff_pion}
Q^2F_{\pi\gamma}(Q^2)=\frac{2A\kappa}{\sqrt{3}\pi}\int_{0}^{1}\frac{dx}{x}\sqrt{x(1-x)}e^{-\frac{1}{2\kappa^2}\left(\frac{m_1^2}{x}+\frac{m_2^2}{1-x}\right)}\left(1-e^{-\frac{(xQ)^2}{2\kappa^2x(1-x)}}\right).
\end{eqnarray}
\end{widetext}
The pion transition form factor is plotted in Figs.~\ref{tff_pion}-\ref{qtff_pion} for $F_{\pi\gamma}(Q^2)$ and $Q^2F_{\pi\gamma}(Q^2)$, both for nonzero and zero quark masses. The pion transition form factor has been measured by BABAR \cite{Aubert2009}, CLEO \cite{Gronberg1998}, CELLO \cite{Behrend1991}, and Belle \cite{Uehara2012}. We compare our model with these experimental data sets up to $Q^2=40$ GeV$^2$ in Figs.~\ref{tff_pion}-\ref{qtff_pion}. The plot for our model with nonzero quark masses fits better with the data. The data for large $Q^2$ values from the BABAR Collaboration do not agree so well with the theoretical prediction from our model, similar to what has been seen in several other studies. However, the data from the Belle Collaboration for the similar range of $Q^2$ values fit better with the model. Note that the form factor with massless quarks fails to fit with the experimental data.

\subsection{$\eta$ and $\eta^\prime$ transition form factors}

\begin{figure}
\begin{center}
\includegraphics[width=\linewidth]{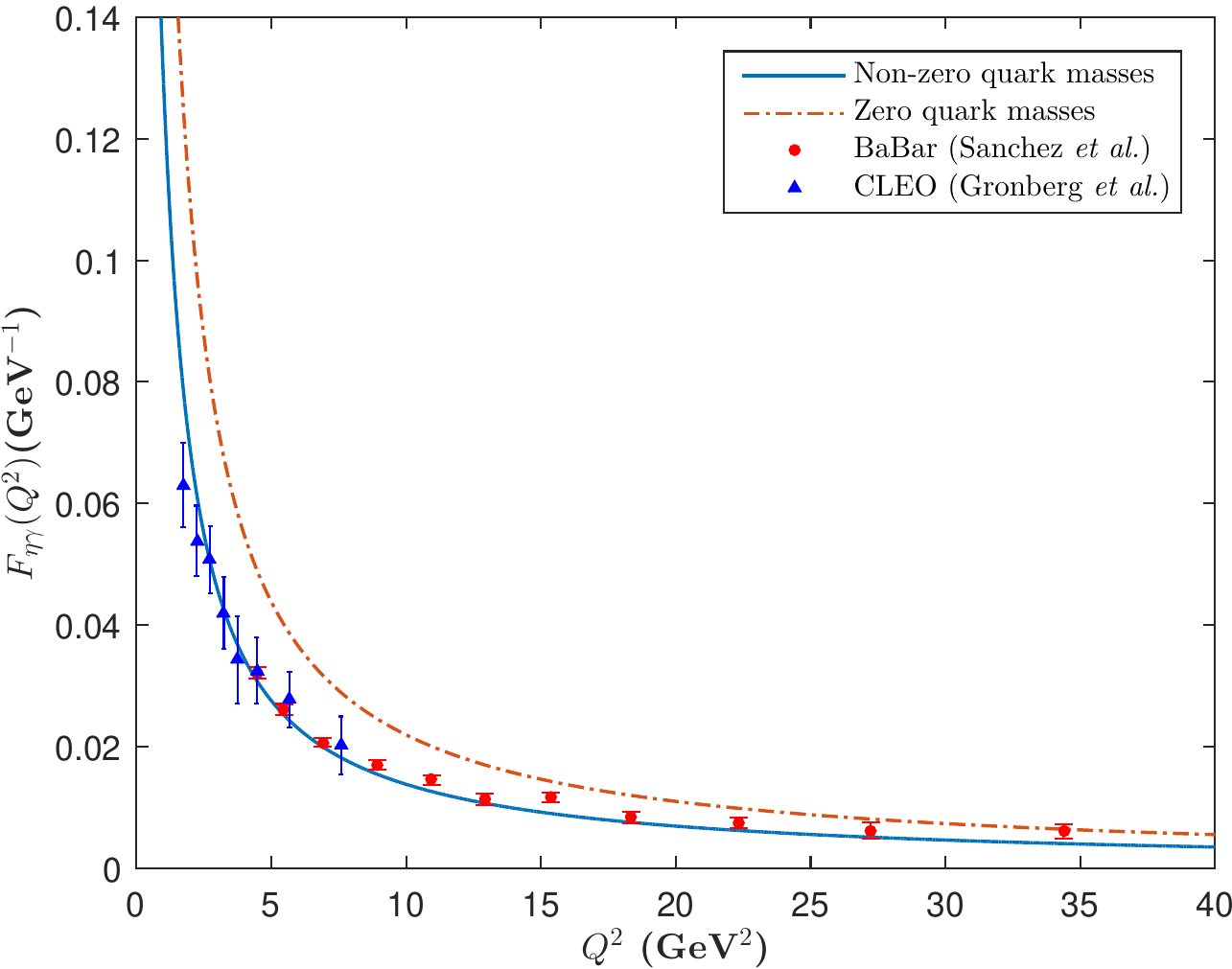} \\
\caption{The $\eta$ transition form factor $F_{\eta\gamma}(Q^2)$ for nonzero (solid line) and zero (dashed line) quark masses. The data are taken from \cite{Sanchez2011,Gronberg1998}.}
\label{tff_eta}
\end{center}
\end{figure}

\begin{figure}
\begin{center}
\includegraphics[width=\linewidth,clip]{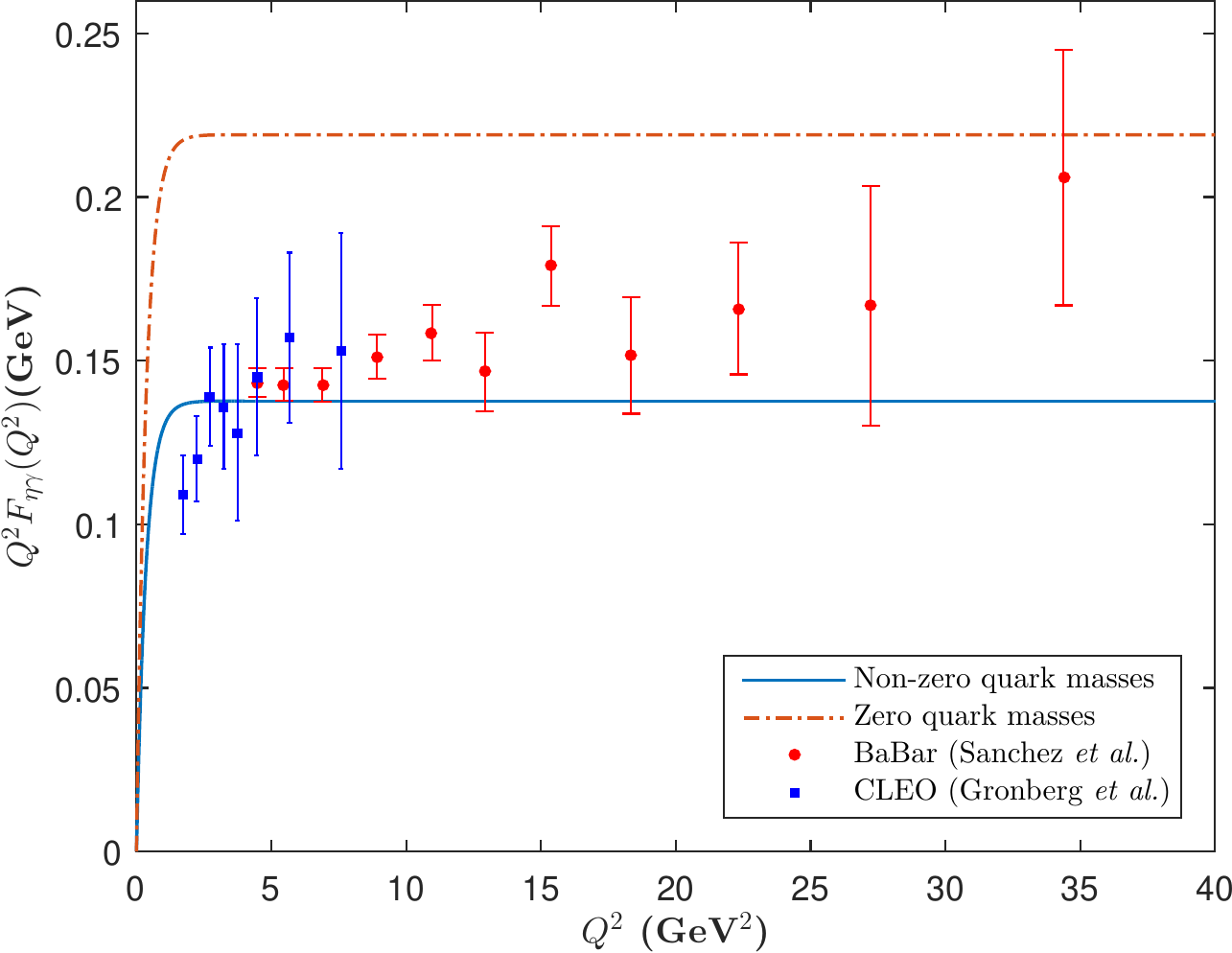} \\
\caption{The $\eta$ transition form factor $Q^2F_{\eta\gamma}(Q^2)$ for nonzero (solid line) and zero (dashed line) quark masses. The data are taken from \cite{Sanchez2011,Gronberg1998}.}
\label{qtff_eta}
\end{center}
\end{figure}

\begin{figure}
\begin{center}
\includegraphics[width=\linewidth,clip]{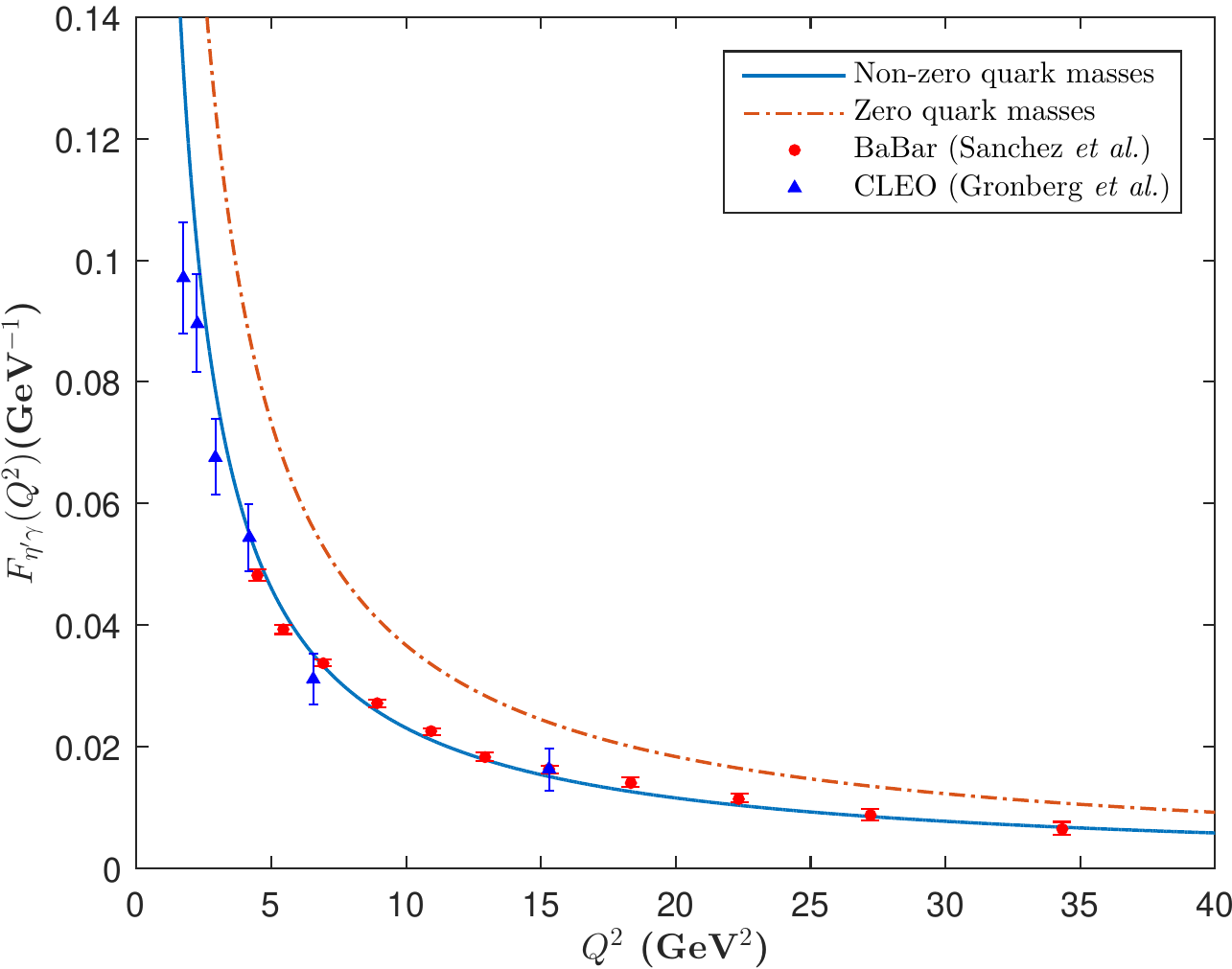} \\
\caption{The $\eta^\prime$ transition form factor $F_{\eta^\prime\gamma}(Q^2)$ for nonzero (solid line) and zero (dashed line) quark masses. The data are taken from \cite{Sanchez2011,Gronberg1998}.}
\label{tff_etap}
\end{center}
\end{figure}

\begin{figure}
\begin{center}
\includegraphics[width=\linewidth,clip]{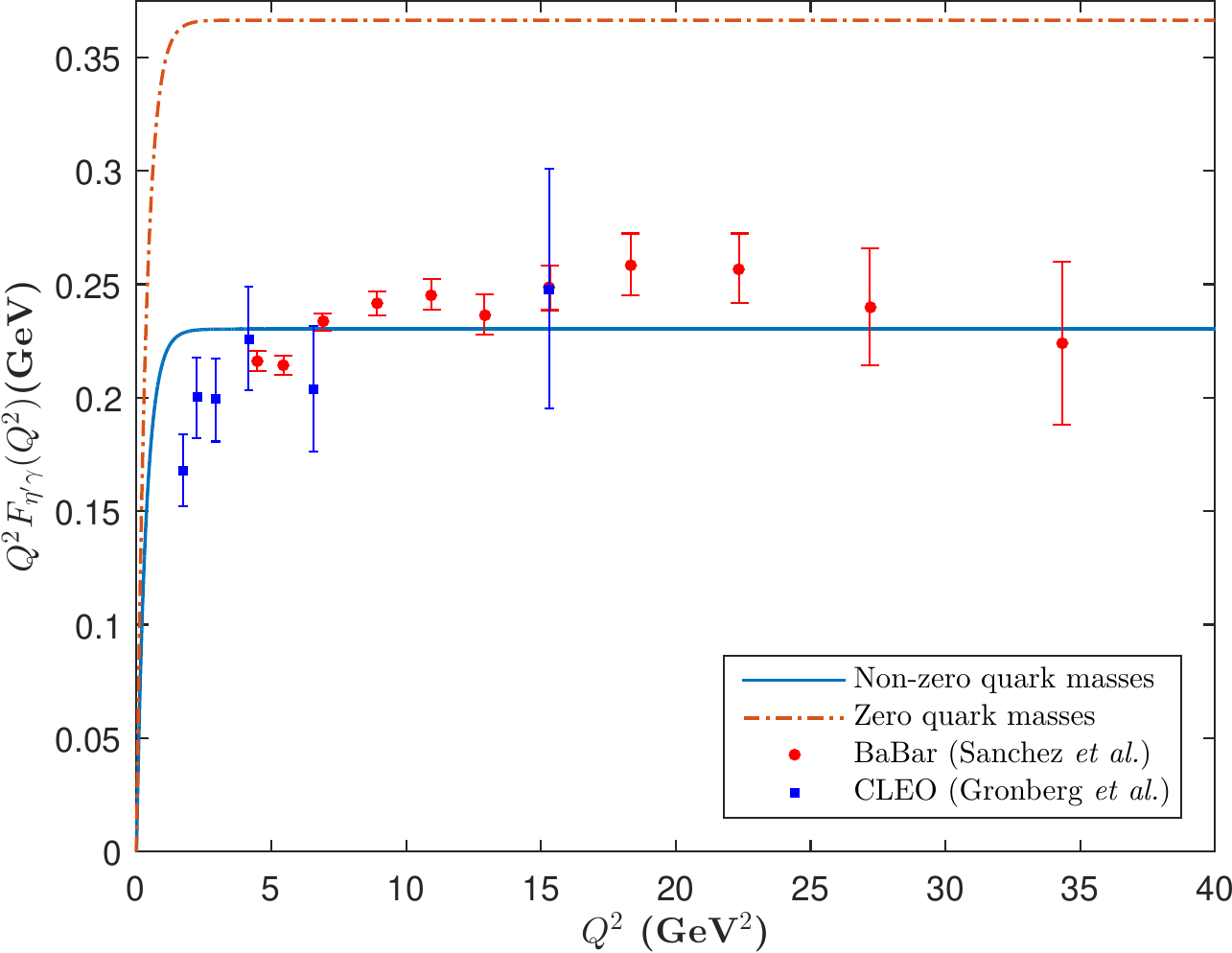} \\
\caption{The $\eta^\prime$ transition form factor $Q^2F_{\eta^\prime\gamma}(Q^2)$ for nonzero (solid line) and zero (dashed line) quark masses. The data are taken from \cite{Sanchez2011,Gronberg1998}.}
\label{qtff_etap}
\end{center}
\end{figure}

The form factors for $\eta$ and $\eta^\prime$ mesons, for the photon-to-meson decay, can be calculated using the prescription used in Ref.~\cite{Brodsky2011}. The $\eta$ and $\eta^\prime$ mesons result from the mixing of neutral states $\eta_8$ and $\eta_1$ given by
\begin{eqnarray}
\eta_8=\frac{u\bar{u}+d\bar{d}-2s\bar{s}}{\sqrt{6}}; \hspace{1mm} \eta_1=\frac{u\bar{u}+d\bar{d}+s\bar{s}}{\sqrt{3}}.\nonumber
\end{eqnarray}
Their transition form factors are given by the same expression as for the pion, except there is an overall constant factor $c_P=\frac{1}{\sqrt{3}}$ and $\frac{2\sqrt{2}}{\sqrt{3}}$ for the $\eta_8$ and $\eta_1$ mesons, respectively. By multiplying $c_P$ with Eq. \eqref{eqtff_pion}, the $\eta_8$ and $\eta_1$ transition form factors are obtained which give the transition form factors for the physical states $\eta$ and $\eta^\prime$ through the following mixing of eigenstates
\begin{eqnarray}   \label{eta_mixing}
\left(\begin{array}{c} F_{\eta\gamma}\\ F_{\eta^\prime\gamma} \end{array}\right) = \left( \begin{array}{cc}
\cos\theta & -\sin\theta \\ \sin\theta & \cos\theta \end{array} \right) \left( \begin{array}{c} F_{\eta_8\gamma} \\ F_{\eta_1\gamma} \end{array} \right).
\end{eqnarray}
The mixing angle is $\theta=-11.4^{\circ}$ \cite{PDG2014}. The plots for $F_{M\gamma}(Q^2)$ and $Q^2F_{M\gamma}(Q^2)$ for $\eta$ and $\eta^\prime$ are shown in Figs.~\ref{tff_eta}-\ref{qtff_eta} and Figs.~\ref{tff_etap}-\ref{qtff_etap}, respectively. These transition form factors have been measured by BABAR \cite{Sanchez2011} and CLEO \cite{Gronberg1998}. We compare our model with these experimental data sets up to $Q^2=40$ GeV$^2$ in Figs.~\ref{tff_eta}-\ref{qtff_etap}. The data from the BABAR Collaboration for $\eta$ and $\eta^\prime$ have better agreement with the model's prediction than in the case of $\pi$. For the transition form factors, the quark masses seem to play a major role and, particularly at large $Q^2$, significant improvements are observed over the model with massless quarks.

\section{Calculation of $\chi^2$ per degree of freedom}
\label{sec_chi}

Before looking at the accuracy of the predictions of this model, it needs to be emphasized that the purpose of this work is to demonstrate the improvement in agreement with the experimental data of the AdS/QCD model for meson, when the wave functions are modified to include the quark masses. The model still represents a valence quark picture and, hence, is not expected to reproduce the experimental data with high accuracy. From Figs.~\ref{ff_pion}-\ref{qtff_etap}, it is evident that the inclusion of quark masses considerably improves the model's predictions. The only tunable parameter in the AdS/QCD model is $\kappa$, which is already fixed by the Regge trajectory, and, hence, we have no free parameter to fit. For the sake of a quantitative comparison, in Table~\ref{chi}, we have listed the values of $\chi^2$ per degree of freedom for different meson properties that we have computed in this paper. Since we have compared two models, the uncertainty in the value of $\kappa$, which is the same for the massive and the massless quark models, is not considered in the $\chi^2$ computation.

From Table~\ref{chi} it can be inferred that, except for the pion and the kaon form factors, the massive quark model gives much better values of $\chi^2$ per degree of freedom for the meson properties. The massive quark model reproduces the overall behavior of the pion form factor much better than the massless one, as can be seen from Fig.~\ref{ff_pion}. However, as shown in the inset of Fig.~\ref{ff_pion}, for very small values of $Q^2$, where we have a large cluster of data points, the massive quark model does not fit well with the data and, hence, $\chi^2/$d.o.f. becomes larger for it than that for the massless case. For the kaon form factor, in the inset of Fig. \ref{ff_kaon}, we have magnified the small $Q^2$ region, which clearly shows the improved result of the massive quark model over the massless one. Here, again, $\chi^2/$d.o.f. comes out to be large, primarily due to three data points with very tiny errors at large $Q^2$ values, where the massive quark model deviates considerably. When we remove these three points from the analysis (remember that we are not fitting any parameter here), the $\chi^2/$d.o.f. value for the massive quark model improves drastically over that for the massless one, as shown in the Table~\ref{chi}.

\begin{table}[ht]
\begin{tabular}{| c | c | c |}
  \hline
Meson & $\chi^2/$d.o.f. & $\chi^2/$d.o.f. \\
 property & (massive quarks) & (massless quarks)\\
  \hline
$F_\pi(Q^2)$ & 73.3715 & 21.5423\\
$F_K(Q^2)$ & 174.4685 & 23.333\\
$F_K(Q^2)$ \footnote{Here, we exclude the three data points at large $Q^2$ values shown in Fig.\ref{ff_kaon}} & 0.819 & 4.4018\\
$F_{\pi\gamma}(Q^2)$ & 6.6594 & 104.4687\\
$F_{\eta\gamma}(Q^2)$ & 2.1186 & 62.8514\\
$F_{\eta^\prime\gamma}(Q^2)$ & 4.0627 & 327.4731\\
  \hline
\end{tabular}
\caption{Value of $\chi^2$ per degree of freedom for different properties of mesons presented in Figs.~\ref{ff_pion}-\ref{qtff_etap}.}\label{chi}
\end{table}


\section{Conclusions}
\label{sec_conclusions}

Light-front holographic QCD maps an effective gravity theory defined in a five-dimensional warped anti--de Sitter spacetime to a semiclassical approximation to strongly coupled QCD quantized on the light-front. It provides a way to model the wave functions of hadrons, analyze their spectrum, and their properties and compare with the experimental data. The soft-wall holographic model has been found particularly useful in computing hadron form factors and transition form factors, which are found to be consistent with the currently available experimental data.

In this paper, we have presented the results for the wave functions, distribution amplitudes, and form factors for $\pi,~\rho,~K$, and $J/\psi$ mesons using the wave functions predicted by the soft-wall holographic model in AdS/QCD. Here, we have used wave functions, modified to include quark masses, which follow from the prescription suggested by Brodsky and de T\'eramond \cite{Brodsky2008a}. The results for pion and kaon form factors are compared with the available experimental data from different experiments. In most cases, AdS/QCD results with zero quark mass fail to agree with the experimental data. The modified AdS/QCD wave functions for massive quarks considerably improve the agreement with the data. The distribution amplitudes and form factors are also compared with pQCD predictions and found to be consistent. In fact, the pQCD predictions for the pion and kaon form factors, given by $F_M(Q^2)=8\pi\alpha_s f^2_{M^+}/Q^2$ at large $Q^2$ values, are in excellent agreement with our AdS/QCD model with massive quarks.

We have also presented the photon-to-meson transition form factors for $\pi$, $\eta$, and $\eta^\prime$, which are reasonably consistent with the available experimental data. The BABAR data \cite{Aubert2009} at large $Q^2$ values for the pion transition form factor are not compatible with our model's results, as has been the case with many other theoretical models, while the data from Belle Collaboration \cite{Uehara2012} have better agreement. The BABAR data \cite{Sanchez2011} for $\eta$ and $\eta^\prime$ are more compatible with the model's results than they are for $\pi$. Again, the results for massless quarks are inconsistent with the data. These conclusions are also supported by the values of $\chi^2$ per degree of freedom, computed for the massive and the massless quark models.

Overall, the results for the form factors and the transition form factors predicted from our model with massive quarks are seen to be more consistent with the experimental data than the results from the model with massless quarks. Inclusion of quark masses can be seen as a good approach to extend the already available AdS/QCD models in order to explain hadronic properties, e.g., generalized parton distributions, nucleon form factors, etc.


\section*{Acknowledgment}

R. S. would like to thank S. J. Brodsky, G. de T\'eramond and F.-G. Cao for providing the data used in their papers and for helpful communication. D. C. would like to thank S. Nedelko for a discussion on the domain model of QCD and its possible mapping to the AdS/QCD model.


\end{document}